# A Unified Framework for Integer Programming Formulation of Graph Matching Problems

Bahram Alidaee[1*], Haibo Wang[2], and Hugh Sloan[1]
1 School of Business Administration, The University of Mississippi, balidaee@bus.olemiss.edu, hsloan@bus.olemiss.edu
2 Sanches School of Business, Texas A&M International University, hwang@tamiu.edu

*Abstract*— Graph theory has been a powerful tool in solving difficult and complex problems arising in all disciplines. In particular, graph matching is a classical problem in pattern analysis with enormous applications. Many graph problems have been formulated as a mathematical program and then solved using exact, heuristic, and/or approximated-guaranteed procedures. On the other hand, graph theory has been a powerful tool in visualizing and understanding complex mathematical programming problems, especially integer programs. Formulating a graph problem as a *natural* integer program (IP) is often a challenging task. However, an IP formulation of the problem has many advantages. Several researchers have noted the need for natural IP formulation of graph theoretic problems. The present study aims to provide a unified framework for IP formulation of graph-matching problems. Although there are many surveys on graph matching problems, none is concerned with IP formulation. This paper is the first to provide a comprehensive IP formulation for such problems. The framework includes a variety of graph optimization problems in the literature. While these problems have been studied by different research communities, however, the framework presented here helps to bring efforts from different disciplines to tackle such diverse and complex problems. We hope the present study can significantly help to simplify some of the difficult problems arising in practice, especially in pattern analysis.

## 1. INTRODUCTION

Graph theory has been a powerful tool in solving difficult and complex problems that arise in different disciplines. In particular, *graph matching* has enormous applications. Recent surveys of graph theoretic problems are presented in [1,2,3,4,5,6,7,8,9]. Most graph optimization problems are NP-hard. All problems discussed in this paper are known to be NP-hard, except for the *graph isomorphism* and the *Golomb Ruler* problems whose complexity statuses are still open, [10,11].

Over the years, many graph and combinatorial problems have been formulated as optimization problems and have been solved by using exact, heuristic, or approximation-guaranteed procedures [12,13,14,15,16]. Similarly, graph theory has been a powerful tool in visualizing and understanding complex optimization problems, (see for example, [17,18,19] for integer programming, [20,21,22,23] for image analysis, [24,25,26] for bioinformatics, and [27] for the social sciences).

Since graph theory and optimization are so closely related, the theoretical and algorithmic developments from one area have been successfully used to address difficult and complex problems in the other discipline. Formulating graph problems as *natural* integer programs (IP) is often a challenging task. However, to tackle such difficulty researchers often use an *indirect* IP formulation of the problem and use it as an upper (lower) bound to design algorithms. A series of *indirect* linear programs for seven graph theoretic problems is presented in [28] where it is noted that "the formulation is indirect in the sense that it captures certain properties of an optimal (integer) solution, yet an integral solution (to the formulation) does not provide an integral solution to the original problem, as opposed to standard linear programming relaxation" (Page 588). A *natural* integer programming formulation of a graph problem has several advantages: (1) it is equivalent to the graph problem; (2) its relaxation often serves as a good upper (lower) bound for the graph problem;

and (3) it is often used in designing approximation-guaranteed and heuristic algorithms. A unified approach based on graph matching and using mathematical programming formulations provides avenues for a better understanding of relationships between these complex problems. In connection to metric labeling problems, Kleinberg and Tardos [29] presented *natural* IP formulations for special cases and gave approximation-guaranteed results. They pointed out that "the difficulty of the general case has to do with the absence of a *natural* integer programming formulation for the problem," [30]. In that spirit, Chekuri, Khanna, Naor, and Zosin [30] presented a *natural* IP formulation for the general case with an approximation-guaranteed algorithm. Several other papers, [30,31,32,33], have pointed out the need for *natural* IP formulations of graph problems. An early IP formulation of a matching problem is given in [34]. Luttamazuzi, Pelsmajer, Shen, and Yang [33] suggested a *natural* IP formulation for three graph optimization problems, including *Bandwidth* minimization. A different *natural* IP formulation of Bandwidth minimization is also introduced in [35]. The first approach to natural IP formulations of a *linear arrangement* problem is given in [36,37]. In [38], a *natural* IP formulation of a graceful labeling problem is discussed. A *natural* IP formulation for multi-layer crossing minimization is proposed in [39]. The *natural* IP formulations for the Golomb Ruler problem are presented in [10,40] and in [41], a *natural* IP formulation for a *contact map problem* is given.

In Part I of this paper, we provide a unified approach based on *graph matching* that allows *natural* integer program formulations for many graph theoretic problems. The relationship between these problems and the quadratic assignment problem (QAP) is pointed out. The approaches include an IP formulation for a variety of problems including the list given below. IP formulations for problems 1-12 are presented in Part II, while IP formulations for problems 13-18 are presented in Part I. For some of these problems, we do not know of any other *natural* IP formulations available in the literature, for some other problems, our formulations are different from those that appear in the literature and provide alternative approaches.

1. Several variations of the traveling salesman problem,
2. Bandwidth problem,
3. Linear arrangement problem,
4. Profile minimization in graphs,
5. Minimum cut linear arrangement,
6. Graph and subgraph isomorphism,
7. Largest common subgraph problem,
8. Maximum subgraph matching,
9. Graph coloring and related problems: graph homomorphism, list coloring, optimum cost chromatic partition, sum coloring, graph labeling, *T*-coloring, maximum independent set and clique, and maximum clique partitioning,
10. Metric labeling variations: embedding a graph in a *d*-dimensional mesh, linear arrangement with *d*-dimensional cost, minimizing storage-time product, and task assignment in distributed computing systems,
11. Golomb ruler problem,
12. Contact map problem,
13. Graph orientation
14. Facility layout problem,
15. Quadratic assignment problem,
16. Fixed spectrum frequency assignment,
17. Interval graph completion,
18. Multi-layer crossing minimization.

**PART I**

2. General Approaches
   *Parameters:*
$G=(V,E)$, an undirected graph,

$G=(V,A)$, a directed graph,

$V$, $(|V|=n)$, set of vertices in a graph $G$,

$E \subseteq V \times V$, set of edges in undirected graph $G$ (allowing possible self-loop),

$A \subseteq V \times V$, set of arcs in digraph $G$ (no self-loop is allowed),

$a = (u,v) \in A$, an arc directed from $u$ toward $v$,

$w_e$ ($w_a$), the weight of an edge $e = \{u,v\} \in E$ (an arc $a = (u,v) \in A$),

$\overline{E}$, set of edges in complement of a graph $G$,

$\overline{A}$, set of edges in complement of a digraph $G$, (i.e., removing all 'arcs' in $A$ and adding an 'edge' between every two nodes where there is no arc in $G$. Thus, if $\{u,v\} \in \overline{A}$ then $(u,v),(v,u) \notin A$),

$N(u) = \{v \in V : \{u,v\} \in E\}$, neighbors of a node $u \in V$ in graph $G$,

$N(u) = \{v \in V : (u,v) \in A\}$, neighbors of a node $u \in V$ in digraph $G$,

$N[u] = N(u) \cup \{u\}$,

$f : V \to V'$, a function mapping $V$ in graph $G$ to $V'$ in graph $G'$,

$c(u,u')$, non-negative cost of assigning a node $u \in V$ to a node $u' \in V'$,

$d(\{u,v\},\{u',v'\})$, the *interaction (or communication)* costs (possibly negative) between two edges,

$d((u,v),(u',v'))$, the *interaction (or communication)* costs (possibly negative) between two arcs.

*Decision variables:*

$x_{uu'} = 1$, if a node $u \in V$ is assigned to a node $u' \in V'$, and 0 otherwise,

$y_{ee'} = 1$, if an edge $e \in E$ is assigned to an edge $e' \in E'$, and 0 otherwise,

$y_{aa'} = 1$, if an arc $a \in A$ is assigned to an arc $a' \in A'$, and 0 otherwise,

$y_{ee'}^\tau = 1$, if an edge $e \in E$ is assigned to an edge $e' \in E'$, and $d(e,e') = \tau$ for a constant $\tau$; and 0 otherwise,

$y_{aa'}^\tau = 1$, if an arc $a \in A$ is assigned to an arc $a' \in A'$, and $d(a,a') = \tau$ for a constant $\tau$; and 0 otherwise,

$K \geq 0$, nonnegative variable (in some cases it is a given constant),

$K_u \geq 0$ ($K_{u'} \geq 0$), for $u \in V$ ($u' \in V'$), nonnegative variable.

With no loss of generality, we let elements of $V$ and $V'$ be integer values, i.e., $V=\{1,...,n\}$ and $V'=\{1,...,n'\}$. Notations $u \in V$ and $u' \in V'$ represent nodes in graph (or digraph) G and $G'$, respectively. Since the nodes are integer values thus for ease of presentation, we may use them as the weight of the nodes. With these notations, for $u,v \in V$ and $u',v' \in V'$, we denote $|u-v|$ and $|u'-v'|$, as absolute values of the difference between the numbers representing the nodes. For each problem, we look for a *matching* (possibly many-to-many), $f : V \to V'$, between nodes of two graphs where some conditions need to be satisfied. The imposed conditions depend on the given problem. For most problems, we define a graph $G'=(V', E')$ in the case of the undirected graph; and a digraph $G'=(V', A')$ in the case of the directed graph, with the '*desirable*' output characteristics, then we find a function $f$ to map nodes of graph $G$ to nodes of the defined graph $G'$. Example 1 illustrates this point.

*Example 1.* Consider the *k*-traveling salesman problem (*k*-TSP). In *k*-TSP, we are given a graph $G$ (directed or undirected) with a distance between edges (arcs in the case of digraph). We look for a minimum distance tour (circuit) of *k* cities [42,43]. When *k*=*n*, we have the TSP (see [43,44], for comprehensive surveys, and [45,46] for classification of integer programming formulations). In *k*-TSP, we define a graph $G'=(V', E')$ in the form of a cycle, Figure 1 without arrows, and in directed graph $G'=(V', A')$ in the form of a circuit, Figure 1, with *k* nodes and distance one between consecutive cities. Then we look for a subset $S \subseteq V$ with $|S|=k$, and a one-to-one function, $f : S \to V'$, that minimizes the total distance on the tour (circuit), i.e., a *subgraph isomorphism* of $G$ to $G'$ that minimizes the total distance traveled. In the case of Bottleneck *k*-

TSP, one looks for a tour (circuit) to minimize the maximum distance traveled between any two cities. As we mentioned, here we have a *subgraph isomorphism (SI)* of $G$ to $G'$. Given two undirected graphs $G=(V, E)$ and $G'=(V', E')$, with $|V|=n \geq |V'|=n'$, a *subgraph isomorphism* of $G$ to $G'$ is a graph $\bar{G}=(S,L)$ with $S \subseteq V$ and $L \subseteq E$ such that $|S|=|V'|$, $|L|=|E'|$ and there exists a one-to-one function $f:V' \to S$ satisfying $\{u',v'\} \in E'$ if and only if $\{f(u'), f(v')\} \in L$. Equivalently, SI asks whether there is a subset of edges and vertices of $G$ that is isomorphic to a smaller graph $G'$. A slightly different *k-TSP* may be defined as follows, called *induced k-TSP*. In that, the subgraph is an *induced subgraph isomorphism (ISI)*. Given two undirected graphs $G=(V, E)$ and $G'=(V', E')$, with $|V|=n \geq |V'|=n'$, an *induced subgraph isomorphism* of $G$ to $G'$ is a graph $\bar{G}=(S,L)$ with $S \subseteq V$ and $L \subseteq E$ such that $|S|=|V'|$, $|L|=|E'|$ and there exists a one-to-one function $f:V' \to S$ satisfying $\{u',v'\} \in E'$ if and only if $\{f(u'), f(v')\} \in E$. Equivalently, the question is whether we can delete vertices from $G$ to obtain a graph isomorphic to $G'$. Note that in SI we do not rule out the possibility of existing $\{u,v\} \in E \setminus L$ such that $f(u')=u$ and $f(v')=v$ for some $\{u',v'\} \in E'$. To illustrate the points, consider graph $G$ in Figure 2, where there are several subgraphs with 4 nodes and 4 edges that each create a *4-TSP*. For example, 1-2-3-4-1 is a subgraph which is a *4-TSP*, while it is not an *induced subgraph*. Another 4 nodes 4 edges subgraph which is *4-TSP* is 1-5-2-3-1 but it is not an *induced subgraph*. However, by deleting node 1 we get an *induced subgraph* 2-3-4-5-2 with 4 nodes and 4 edges which creates an *induced 4-TSP*. Integer programming formulation of subgraph isomorphism and induced subgraph isomorphism are presented in Part II.

While in some problems, we look for a *feasible* solution, in other problems, we may need to optimize an objective function. The objective functions are in one of the forms P1-P7 given in Table 1. In the table, objective functions are in a *minimization* format. However, they need to be appropriately modified when formulating a specific problem. Objective functions P1-P7 cover a wide range of combinatorial problems. In the next section, we present several general matching problems where the focus is on undirected graphs, but the formulas can easily be modified for directed graphs.

Based on the aforementioned discussion, we present three general but related problems, outputs 1-3. Output 1 is a feasibility problem. Both outputs 2 and 3, optimize an objective function while satisfying 'some form of feasibility' introduced by output 1. The feasibility in output 1 ensures that nodes of two graphs are assigned to each other while the cost of communication between edges of the two graphs must be limited to a pre-specified set of numbers. Output 2 optimizes one of the objective functions P1-P7 while feasibility in output 1 is satisfied. Output 3 optimizes the sum of penalties for violating the feasibility conditions of output 1. For each output, a natural IP formulation is presented. In each case, proof is available. Due to saving space, we do not present the proof here, however, upon request the proof is available from the authors. These outputs provide a general framework within which many problems including those listed in 1-18 with appropriate modifications of the formulations can be considered as special cases of one of these three outputs. IP formulations for problems 1-12 are presented in Part II, while IP formulations for problems 13-18 are presented in Part I.

-----------------------------
Figure 1 about here
-----------------------------
-----------------------------
Figure 2 here
-----------------------------
-----------------------------
Table 1 about here
-----------------------------

2.1 Three General Matching Problems

Input: Undirected graphs $G=(V, E)$, and $G'=(V', E')$ with possible self-loop for each vertex in $V'$. A set of *allowable* assignments, $L_u \in V'$, for each $u \in V$. A set of integers, $t_e$, (possibly empty) for each edge $e \in E$, called the *forbidden set*. A penalty $p_e^\tau$ associated with each element $\tau \in t_e$. A cost of $c(u,u')$ for assigning a node $u \in V$ to a node $u' \in V'$. A cost of $d(e,e')$ for assigning an edge $e \in E$ to an edge $e' \in E'$. In each output 1-3 below, the task is to find a function $f: V \to V'$ such that $f(u) \in L_u$ for $u \in V$, and $\{f(u), f(v)\} \in E'$ whenever $\{u,v\} \in E$, and

Output 1: $d(\{u,v,\},\{f(u),f(v)\}) \notin t_{\{u,v\}}$.

Output 2: $d(\{u,v,\},\{f(u),f(v)\}) \notin t_{\{u,v\}}$, to minimize one of the objectives P1-P7.

Output 3: if $d(\{u,v,\},\{f(u),f(v)\}) = \tau \in t_{\{u,v\}}$ there is a penalty $p_{\{u,v\}}^\tau$. The task is to minimize the total penalty.

In the following, we will present the IP formulation for each output 1-3. We will point out problems from the list that are special cases of the outputs. However, the IP formulations for outputs 1-3 need to be appropriately modified to represent these special cases, presented in Part II.

### 2.1.1 Formulations for Output 1
The matching problem with output 1 is known as the T-coloring problem [47,48]. An IP formulation of the output 1 is as follows.

$$\sum_{u' \in L_u} x_{uu'} = 1, \text{ for } u \in V, \tag{1}$$

$$x_{uu'} + x_{vv'} \le 1, x_{uv'} + x_{vu'} \le 1, \text{ for } \{u,v\} \in E, \text{ and } \{u',v'\} \in \overline{E}', u' \ne v', \tag{2}$$

$$x_{uu'} + x_{vv'} \le 1, x_{uv'} + x_{vu'} \le 1, \text{ for } \{u,v\} \in E, \text{ and } \{u',v'\} \in E', \text{ and } d(\{u,v\},\{u',v'\}) \in t_{\{u,v\}}. \tag{3}$$

Constraints (1) ensure node assignments are from allowable sets. Constraints (2) ensure that the endpoints of each edge $\{u,v\} \in E$ can only be assigned to end points of an edge from $E'$. Here, for a node in graph $G'$ self-loop is allowed. Thus, endpoints of an edge in $E$ can be assigned to one node in $V'$. Constraints (3) forbid an edge $e$ in $E$ to be assigned to an edge $e'$ in $E'$ if communication cost $d(e,e')$ is in the set $t_e$.

Modification of these formulations for directed graphs is given below. In that, constraint (1) has the same effect as was explained. Since a complement to a directed graph is a graph with edges, thus constraints (4) include edges instead of arcs in the complement of graph $A'$. Constraints (4) ensure that the endpoints of each arc $(u,v) \in A$ can only be assigned to the endpoints of an arc from $A'$. Constraints (5) ensure that arcs are correctly assigned to each other. In that, if two arcs $(u,v)$ and $(u',v')$ are assigned to each other, we must have $u'=f(u)$ and $v'=f(v)$. Constraints (6) forbid two arcs $a$ and $a'$ to be assigned to each other if the communication cost $d(a, a')$ is in the set $t_a$.

$$\sum_{u' \in L_u} x_{uu'} = 1, \text{ for } u \in V, \tag{1}$$

$$x_{uu'} + x_{vv'} \le 1, x_{uv'} + x_{vu'} \le 1, \text{ for } (u,v) \in A, \text{ and } \{u',v'\} \in \overline{A}', \tag{4}$$

$$x_{uv'} + x_{vu'} \le 1, \text{ for } (u,v) \in A, \text{ and } (u',v') \in A', \tag{5}$$

$$x_{uu'} + x_{vv'} \le 1, \text{ for } (u,v) \in A, \text{ and } (u',v') \in A', \text{ and } d((u,v),(u',v')) \in t_{(u,v)}. \tag{6}$$

### 2.1.2 Formulations for Output 2
*2.1.2.1. Output 2 and objective function P1:*

The matching problem with output 2 and the objective function P1 is a bottleneck minimization that includes as a special case: Bottleneck *k*-TSP, bottleneck *induced k-TSP*, Bandwidth minimization, several variations of graph coloring, graph labeling, and the Golomb Ruler problems. An IP formulation of the problem can be stated as follows. Note that the formulations need to be appropriately modified for special cases. Formulations for these special cases are presented in Part II.

*Min K,*   s.t. (1-3), $K \geq 0$, and

$$x_{uu'} + x_{vv'} \leq y_{\{u,v\}\{u',v'\}} + 1, \quad x_{uv'} + x_{vu'} \leq y_{\{u,v\}\{u',v'\}} + 1, \text{ for } \{u,v\} \in E, \text{ and } \{u',v'\} \in E', \tag{7}$$

$$\sum_{t' \in L_u} c(u,u') x_{uu'} + \sum_{t' \in L_u} \sum_{t' \in L_v \cap N[u']} d(\{u,v\}\{u',v'\}) y_{\{u,v\}\{u',v'\}} \leq K, \text{ for } u \in V, v \in N(u). \tag{8}$$

Constraints (1-3) ensure the feasibility of output 1 is satisfied. Constraints (7) ensure that $y_{ee'} = 1$ if and only if two edges $e$ and $e'$ are assigned to each other which adds a value of $d(e,e')$ to the LHS of (8). If a node $u \in E$ is assigned to a node $u' \in E'$, and if a node $v \in N(u)$ is assigned to $v' \in N[u']$, i.e., a neighboring edge of $u$ assigned to a neighbor edge of $u'$ then it can add a cost of $c(u,u')+d(\{u,v\},\{u',v'\})$ to the objective function. Since we minimize $K$, the largest cost over all nodes $u \in V$ is being minimized.
Example 2 illustrates the formulation when used in bottleneck $k$-TSP and bottleneck-induced $k$-TSP.

*Example 2.* Given graph $G=(V, E)$, the problem is to find (1) an *induced k-TSP*, and (2) a *k-TSP*, to minimize the longest distance between any two cities, where $w_{\{u,v\}}$ is a given real number representing the distance between two cities $u$ and $v$.
To give IP formulation, we create a graph $G'=(V', E')$ with $|V'|=k$ nodes similar to the graph in Figure 1 without arrows. Then we want to find: (1) an ISI of $G$ and $G'$ to minimize the longest distance between every two cities, and (2) an SI of $G$ and $G'$ to minimize the longest distance between every two cities. Let $c(u,u')=0$ for all $u \in V$ and $u' \in V'$. Also, let the distance between every two cities with an edge in $G'$ be equal to one. Define $d(e,e') = w_e w_{e'}$ for all $e \in E$ and $e' \in E'$. To give IP for *induced k-TSP*, we need to find subsets $S \subseteq V$ and $L \subseteq E$ such that $|S|=k$, and $|L|=|E'|$, and a one-to-one function $f: S \to V'$ satisfying $\{u',v'\} \in E'$ if and only if $\{f(u'), f(v')\} \in E$ to minimize the longest distance between every two cities in $S$. Here, clearly, the objective function is a special case of P1, as follows.

*P*1:   $Min_f \{Max_{u \in V, v \in N(u)} \{d(\{u,v\}, \{f(u), f(v)\})\}\} = Min_f \{Max_{e \in E, e' \in E'} \{d(e,e')\}\}.$

A modification of the above IP for this special case is as follows. We need constraints (9) to create a subset $S$ and a one-to-one function between nodes of $V'$ and $S$.

$$\sum_{u' \in V'} x_{uu'} \leq 1, \text{ for } u \in V, \qquad \sum_{u \in V} x_{uu'} = 1, \text{ for } u' \in V'. \tag{9}$$

Furthermore, to have *induced subgraph isomorphism* we need constraints (10-11).

$$x_{uu'} + x_{vv'} \leq 1, \quad x_{uv'} + x_{vu'} \leq 1, \text{ for } \{u,v\} \in E, \text{ and } \{u',v'\} \in \overline{E}', \tag{10}$$

$$x_{u'u} + x_{v'v} \leq 1, \quad x_{v'u} + x_{u'v} \leq 1, \text{ for } \{u,v\} \in \overline{E}, \text{ and } \{u',v'\} \in E'. \tag{11}$$

If an edge $\{u,v\}$ is assigned to an edge $\{u',v'\}$ then we have $y_{\{u,v\}\{u',v'\}} = 1$ ensured by constraints (7).

$$x_{uu'} + x_{vv'} \leq y_{\{u,v\}\{u',v'\}} + 1, \quad x_{uv'} + x_{vu'} \leq y_{\{u,v\}\{u',v'\}} + 1, \text{ for } \{u,v\} \in E, \text{ and } \{u',v'\} \in E'. \tag{7}$$

For a given node $u \in V$, consider a neighbor node $v \in V$ (i.e., there is an edge $e = \{u,v\} \in E$). If the edge $e$ is assigned to an edge $e'=\{u',v'\}$ then we have $d(e,e') = w_e w_{e'} = w_e$. The objective is to minimize the maximum of $d(e,e')$ for all $e \in E$ and $e' \in E'$ when assigned to each other, this is done by adding an objective function, *Min K*, and constraints (12).

$$d(\{u,v\}\{u',v'\}) y_{\{u,v\}\{u',v'\}} \leq K, \quad d(\{u,v\}\{v',u'\}) y_{\{u,v\}\{u',v'\}} \leq K, \text{ for } \{u,v\} \in E, \text{ and } \{u',v'\} \in E'. \tag{12}$$

Constraints (12) can also be replaced by constraints (13) that are the same as constraints (8). Note that in (13), given an edge *{u,v}*, only if it is assigned to an edge *{u',v'}* then we have $y_{\{u,v\}\{u',v'\}} = 1$. Here, given a node *u* and one of its neighbors $v \in N(u)$ (i.e., there is an edge $e = \{u,v\} \in E$), since we already have ensured a one-to-one correspondence between *S* and *V'* by (9) then the left-hand side (LHS) of the (13) will be exactly equal to the LHS of (12).

$$\sum_{u' \in L_u} \sum_{v' \in L_v \cap N[u']} d(\{u,v\}\{u',v'\}) y_{\{u,v\}\{u',v'\}} \leq K, \quad \text{for } u \in V, v \in N(u). \tag{13}$$

Now, to appropriately modify the formulations for *k-TSP*, we only need to remove the constraints (10) from the set of constraints presented for *induced k-TSP*. Note that, different but simpler formulations for *k-TSP* and induced *k-TSP* are presented in Part II.

*2.1.2.2. Output 2 and objective function P2:*

The matching problem with output 2 and objective function P2 includes as a special case: *k*-TSP, *induced k-TSP*, variations of coloring problems, linear arrangement, variations of metric labeling, the largest common subgraph problem, maximum subgraph matching, facility layout, and contact map problem. An IP formulation is as follows. Similar to the case for P1, the formulations for special cases need to be appropriately modified. They are presented in Part II.

$$\text{Min } \sum_{u \in V} \sum_{u' \in L_u} c(u,u') x_{uu'} + \frac{1}{2} \sum_{u \in V} \sum_{v \in N(u)} \sum_{u' \in L_u} \sum_{v' \in L_v \cap N[u']} d(\{u,v\}\{u',v'\}) y_{\{u,v\}\{u',v'\}} \tag{14}$$

s.t. (1-3),(7).

Given a matching between the two graphs, the objective function P2 sums up all node assignments and all edge assignments costs and minimizes the total quantity, constraints (1-3) ensure the feasibility of assignments. Constraints (7) ensure that communication costs between edges are added to the total cost. Note that the objective function (14) has a similarity with constraints (8). In constraints (8), for a given node *u* and one of its neighbors *v* (i.e., $e = \{u,v\} \in E$), if assigned respectively to *u'* and *v'* (i.e., $e' = \{u',v'\} \in E'$), then we considered the total cost of *c(u,u')+d(e,e')*, and minimize largest of such cost. However, in (14) the sum of all such costs is minimized. Example 3 illustrates the formulation when used in *k-TSP* and *induced k-TSP*.

*Example 3*. The problem is similar to the problem of example 2, however, the objective is the minimization of the total distance traveled.
Here, parameters are defined similar to example 2. However, we minimize the following quantity for all *e* and *e'* assigned to each other.

$$\sum_{e \in E} \sum_{e' \in E'} d(e,e'). \tag{15}$$

This is the same as P2 given as follows.

$$P2: \quad Min_f \left\{ \sum_{\{u,v\} \in E} d(\{u,v\},\{f(u),f(v)\}) \right\}.$$

An IP formulation for *induced k-TSP* is as follows.

$$\text{Min } \sum_{e \in E} \sum_{e' \in E'} d(e,e') y_{ee'} = \frac{1}{2} \sum_{u \in V} \sum_{v \in N(u)} \sum_{u' \in V'} \sum_{v' \in N(u')} d(\{u,v\}\{u',v'\}) y_{\{u,v\}\{u',v'\}} \tag{14}$$

s.t. (7), (9-11).

An IP formulation for *k-TSP* is similar except that constraints (10) are removed. Note that, here also different but simpler formulations for *k-TSP* and induced *k-TSP* are presented in Part II.

*2.1.2.3. Output 2 and objective function P3:*

The matching problem with output 2 and objective function P3 includes the profile minimization problem (PMP) as a special case. The PMP is equivalent to the interval graph completion problem, which is to find a super-graph of a graph $G$ with as small a number of edges as possible. Given an undirected graph $G=(V,E)$, the PMP looks for a one-to-one function $f:V \to \{1,\ldots,|V|\}$ such that $\sum_{u \in V} Max_{v \in N[u]}\{f(u)-f(v)\}$ is minimized. The objective function in PMP is a special case of P3. Example 4 below illustrates this point.

*Example 4.* Define a complete undirected $G'=(V',E')$ with $|V'|=\{1,\ldots,|V|\}$. For each $u \in V$ let $c(u,f(u))=f(u)$, and for each $\{u,v\} \in E$ let $d(\{u,v\},\{f(u),f(v)\})=-Min[f(u),f(v)]$. The objective of PMP is now equal to P3 given as

P3: $Min_f \left\{ \sum_{u \in V} Max_{v \in N(u)} \left\{ c(u,f(u)) + d(\{u,v\},\{f(u),f(v)\}) \right\} \right\}$.

IP formulation for PMP is presented in Part II. An IP formulation of output 2 with the objective function P3 is given below. Note that the objective function P3 has similarity with P1. Given a node $u \in V$ the maximization part of P3 (the quantity in the summation) is the same as the maximization in P1. However, P3 sums up such quantity over all nodes $u$.

$Min \sum_{u \in V} K_u$, s.t. (1-3), (7), $K_u \geq 0$, and

$$\sum_{u' \in L_u} c(u,u')x_{uu'} + \sum_{u' \in L_u} \sum_{v' \in L_v \cap N[u']} d(\{u,v\}\{u',v'\})y_{\{u,v\}\{u',v'\}} \leq K_u, \text{ for } u \in V, v \in N(u). \quad (16)$$

Constraints (1-3) and (7) have the same effect as was explained earlier. For a given node $u \in V$, the LHS in (16) is the same as constraints (8), however, the objective function minimizes the sum of all such quantities.

### 2.1.2.4. Output 2 and objective function P4:

The matching problem with output 2 and the objective function P4 includes several variations of metric labeling, and task assignment in distributed computing systems (TADCS) as special cases. Here, we are looking for a function (not necessarily one-to-one) $f:V \to V'$ between nodes of two graphs. If a node $u \in V$ is assigned to a node $u' \in V'$ there is a cost of $c(u,u')$. Given a node $u'$ there is a possibility that several nodes from $V$ are assigned to it. If an edge $e \in E$ is assigned to an edge $e' \in E'$ then there is a cost of $d(e,e')$. Thus, the total cost of assigning all nodes $u \in V$ to a given node $u' \in V'$ is equal to

$$\sum_{u \in V: f(u)=u'} c(u,f(u)) + \sum_{\substack{\{u,v\} \in E:(f(u)=u' \text{ or } f(v)=u')\\ \text{and } \{f(u),f(v)\} \in E'}} d(\{u,v\},\{f(u),f(v)\}). \quad (17)$$

The objective is to find a matching function that minimizes the maximum of (17) over all nodes $u' \in V'$ which is exactly P4. Given a node $u'$, the LHS in (18) is the total cost equivalent to (17). The reason that the triple summation is divided by 2 is that we are assuming if $u' \in L_u$ then $u' \in L_v$ for $v \in N(u)$. If this assumption is not satisfied the formulation needs some minor modification. Thus, an IP formulation is as follows.

$Min\ K$, s.t. (1-3),(7), $K \geq 0$, and

$$\sum_{u:u' \in L_u} c(u,u')x_{uu'} + \frac{1}{2} \sum_{u:u' \in L_u} \sum_{v \in N(u)} \sum_{v' \in L_v \cap N[u']} d(\{u,v\}\{u',v'\})y_{\{u,v\}\{u',v'\}} \leq K, \text{ for } u' \in V'. \quad (18)$$

A special case of the above formulation under the *task assignment in distributed computing systems (TADCS)* is discussed in [49]. Example 5 illustrates this case.

*Example 5.* In the TSDCS, we have a set of tasks, $V$, and a set of processors, $V'$. A *task graph* $G=(V, E)$ represents the set of tasks and their communication links shown by the edges of the graph. A *processor graph* $G'=(V', E')$ represents the set of processors and their communication links shown by the edges of the graph. The problem is to find an assignment function $f : V \to V'$ that assigns the tasks to the processors. If the task $u \in V$ is assigned to an eligible processor $u' \in V'$, the processing time is $c(u,u')$. If two tasks $u, v \in V$ need to communicate (i.e. if there is an edge $\{u,v\} \in E$), they can be assigned to one processor; in that case, there is no communication time between them. In this case, we are assuming node $u'$ has a self-loop. Two tasks can also be assigned to two different processors if a communication link is possible between them (i.e. if there is an edge $\{u',v'\} \in E'$). In that case, there is a communication time between two tasks measured by $d(\{u,v\},\{f(u),f(v)\})$. The problem asks for an assignment $f : V \to V'$ to minimize the completion time of the last task completed, i.e., the load of the processor with the maximum processing and communication times [49], which is a special case of P4. However, if the problem is to minimize total times (see for example, [50,51,52,53] for special cases of such scheduling problems) then we have the objective P2 to be minimized. Several other specific examples of metric labeling with objective functions P2 or P4 are presented in Part II. Prominent applications are in pattern analysis, and image processing, (see for example, [28,29,30]).

*2.1.2.5. Output 2 and objective function P5:*

The matching problem with output 2 and the objective function P5 includes several variations of graph coloring as special cases including a problem of transmission of real-time messages in a metropolitan network or a problem related to dynamic storage allocation, [54,55], and scheduling on a batch machine with job compatibilities, [55,56,57]. In P5, if a node $u$ is assigned to a node $u'$ the total costs are a sum of node assignment cost, $c(u,u')$, and edge assignment costs. It is the sum of $d(\{u,v\},\{f(u),f(v)\})$ for all $v$ such that $\{u,v\} \in E$ and $\{f(u), f(v)\} \in E'$ which is equal to

$$c(u, f(u)) + \sum_{v \in N(u)} d(\{u,v\}, \{f(u), f(v)\}). \tag{19}$$

Among all nodes, $u$ assigned to a given node $u'$ we are looking at the maximum that is equal to

$$Max_{u \in V : f(u) = u'} \left\{ c(u, f(u)) + \sum_{v \in N(u)} d\left(\{u,v\}, \{f(u), f(v)\}\right) \right\}. \tag{20}$$

The problem looks for a function $f : V \to V'$ to minimize the sum of all such maximum costs over all nodes $u' \in V'$ which is exactly P5. An IP formulation of the problem can be stated as follows.

$Min \sum_{u' \in V'} K_{u'}$, s.t. (1-3),(7), $K_{u'} \geq 0$, for $u' \in V'$, and

$$c(u,u')x_{uu'} + \sum_{v \in N(u), u' \in L_v} d(\{u,v\}\{u',v'\})y_{\{u,v\}\{u',v'\}} \leq K_{u'}, \text{ for } u' \in V', \text{ and } u \in V \text{ with } u' \in L_u. \tag{21}$$

The LHS in (21) is the cost contributed by node $u$ when assigned to a node $u'$ plus the cost communicated by each edge neighboring $u$ assigned to each edge neighboring $u'$. Since the RHS is the same for all $u$ thus the objective function P5 minimizes the sum over all such values. Also, note that constraints (1-3) and (7) play the same role as explained before.
Example 6 illustrates a task assignment problem as a special case of the above problem.

*Example 6*. The problem is similar to the TSDCS defined in example 5. However, for a given processor $u'$ the total cost is defined by (19). Special cases of this problem are scheduling on a batch machine with job compatibilities where communication costs are zero, [55,56,57]. In this special case, the problem becomes a *Weighted Vertex Coloring Problem (WVCP)* discussed in more detail in Part II. Note that in TSDCS since a task $u \in V$ must communicate with its neighbors, i.e., with each $v \in N(u)$, thus if $u$ is assigned to $u'$ and if task $v$ is assigned to the processor $u'$ there are no communication costs however if it is assigned to a processor $v' \in V'$ with $u' \neq v'$ then the communication cost is $d(\{u,v\},\{u',v'\})$. Thus, the total cost of assigning a task $u \in V$ to a given processor $u' \in V'$ is equal to the LHS of (20). The value of $K_{u'}$ is the largest overall tasks $u \in V$, where the sum is minimized.

*2.1.2.6. Output 2 and objective function P6:*

The matching problem with output 2 and the objective function P6 generalizes the *graph orientation (GO)* problem [58,59,60], and the edge covering problem [61]. Other names for GO are *edge linking* [62], *link orientation* [63], and *vertex orderings* [64].

In GO, the problem is to determine a balanced ordering of the vertices of a graph; that is, the neighbors of each vertex *u* are as evenly distributed to the left and right of *u* as possible. The problem has applications in a variety of areas including in graph drawing [64] and providing quality of service (QoS) networks [60,63]. The major objectives of various information services on modern networks are to provide high efficiency, good quality, and maximum throughput for the system resources [60,63]. The GO has been proposed for assigning flow orientations over links in a network. Refer to [60,63] for applications, proof of NP-hardness, and several polynomial solvable special cases. Refer to [61] for a recent IP formulation. Sadig, Mozafari, and Kashan [61] mention other applications including the edge covering (EC) problem. The EC problem has applications in locating emergency facilities such as police stations, and road hospitals or an electronic network for locating information centers. The main characteristic of these applications is that the customer's demand is distributed uniformly in the paths between the location points [61].

The GO can be defined as follows, [60,63]. Given an undirected graph *G=(V, E)*, each node *u* is associated with a weight (cost) $\alpha_u$, and each edge $\{u,v\} \in E$ is associated with two weights (costs) $w_{(u,v)}$ and $w_{(v,u)}$ where $(u,v)$ is directed from *u* to *v* (an orientation from *u* to *v*, or an out-degree node of *u*) and $(v,u)$ is directed from *v* to *u* (an orientation from *v* to *u*, or an in-degree node of *u*). Thus, each edge is considered as two directed arcs. Given an orientation for all edges, for a node $u \in V$ defined out-degree nodes, *ON(u),* and in-degree nodes, *IN(u)*, as follows

$ON(u) = \{v : (u,v) \text{ is an out-degree node of } u\}$,

$IN(u) = \{v : (v,u) \text{ is an in-degree node of } u\}$.

The problem is to find an orientation for each edge. Given an orientation for a graph, the total cost associated with each node is equal to $\alpha_u + \sum_{v \in ON(u)} w_{(u,v)}$. The GO looks to choose an orientation for each edge to minimize the largest costs for all node $u \in V$ which is the minimization of $Max_{u \in V} \{\alpha_u + \sum_{v \in ON(u)} w_{(u,v)}\}$.

Output 2 with objective function P6, generalizes the GO problem and the EC problem, which can be stated as follows. We are given two undirected graphs *G=(V,E)* and *G'=(V',E')*, we want to find a function $f : V \to V'$. In that, if a node *u* is assigned to a node *u'* the cost is the same as (19). Among all nodes $u \in V$ we are looking at the maximum cost which is equal to

$$Max_{u \in V} \left\{ c(u, f(u)) + \sum_{v \in N(u)} d\left(\{u,v\}, \{f(u), f(v)\}\right) \right\}. \tag{22}$$

Note that, in (22) for simplicity of presentation we did not include edge orientation, i.e., choosing a direction for each edge. To consider edge orientation minor modification in the formulation is needed. When edge orientation needs to be considered value of *d({u,v},{f(u),f(v)})* for *{u,v}* depends on the direction of the edge, i.e., we have
$d(\{u,v\}, \{f(u), f(v)\}) \neq d(\{v,u\}, \{f(v), f(u)\})$.

The objective is to fund a function *f* to minimize (22) which is exactly P6. Now, an IP formulation of the problem can be stated as follows:

*Min K*,    s.t. (1-3), (7), $K \geq 0$, and

$$\sum_{u' \in L_u} c(u, u') x_{uu'} + \sum_{u' \in L_u} \sum_{v \in N(u)} \sum_{v' \in L_v \cap N[u']} d(\{u,v\}\{u',v'\}) y_{\{u,v\}\{u',v'\}} \leq K, \text{ for } u \in V. \tag{23}$$

Constraints (1-3) and (7) have the same effect as we described before. The LHS in (23) is representing the cost of assigning a given node $u \in V$ to a node $u' \in V'$ plus the cost of all edges neighboring *u,* if assigned to edges in *E'*. The objective function P6 minimizes the largest of such quantity.

### 2.1.2.7. Output 2 and objective function P7:

The matching problem with output 2 and the objective function P7 includes the *Minimum Cut Linear Arrangement Problem (MCLAP)* as a special case (see [6] for recent discussion and complexity of special cases). Given an undirected graph *G=(V, E)*, MCLAP looks for the smallest value of *K* and a one-to-one function $f : V \to \{1,\ldots,n\}$ such that for all $1 \leq i \leq n$, $|\{\{u,v\} \in E, f(u) \leq i < f(v)\}| \leq K$. IP formulation for MCLAP is presented in Part II. It has applications in telecommunication [65], computer-aided design, automatic graph drawing, protein engineering, optimization of rail systems [2], and computational logic [66], among others.

In output 2 with objective function P7, we are looking for a function $f : V \to V'$. In that, given a node $i' \in V'$, if endpoints of an edge *{u,v}* in *E* are assigned to end points of an edge *{f(u),f(v)}* in *E'* where $f(u) \leq i' < f(v)$ then the cost is

$$c(u, f(u)) + c(v, f(v)) + d(\{u,v\}, \{f(u), f(v)\}).$$

We want to minimize the maximum of such quantity over all nodes $i' \in V'$. An IP formulation can be stated as follows:

*Min K,* s.t. (1-3),(7), $K \geq 0$, and

$$\sum_{\substack{\{u,v\} \in E, \{u',v'\} \in E', \\ f(u)=u' \leq i' < f(v)=v'}} (c(u, u') + c(v, v') + d(\{u,v\}, \{u', v'\})) y_{\{u,v\}\{u',v'\}} \leq K, \text{ for } i' \in V'. \tag{24}$$

Constraints (1-3) and (7) have the same effects as described before. The left-hand side of (24) represents the cost contributed by all edges $\{u,v\} \in E$ if *u* and *v* in *V* are respectively assigned to two nodes *u'* and *v'* in *V'* where $\{u',v'\} \in E'$ and such that $u' \leq i' < v'$ for a given node $i' \in V'$. The objective function minimizes this cost over all nodes $i' \in V'$.

### 2.1.3 Formulations for Output 3

The matching problem with output 3 is a generalization of P2. In that, we allow edges to be assigned to each other if the cost of an assignment is in the *forbidden set*. However, this is done at the expense of a penalty, and the task is to minimize the total penalty. It includes the *Fixed Spectrum Frequency Assignment* (FSFA), [67,68,69,70], as a special case. Montemanni, Smith, and Allen in [68,70] presented an IP formulation of FSFA; however, our formulation given below for the general case when applied to FSFA is different and is simpler with fewer variables.

*Min* $\sum_{u \in V} \sum_{v \in N(u)} \sum_{u' \in L_u} \sum_{v' \in L_v \cap N[u']} \sum_{\tau \in t_{\{u,v\}}} p^{\tau}_{\{u,v\}} y^{\tau}_{\{u,v\}\{u',v'\}}$

s.t. (1-2), and $x_{uu'} + x_{vv'} \leq y^{\tau}_{\{u,v\}\{u',v'\}} + 1$, $x_{uv'} + x_{vu'} \leq y^{\tau}_{\{u,v\}\{u',v'\}} + 1$,

for $\{u,v\} \in E, \{u',v'\} \in E', d(\{u,v\},\{u',v'\}) = \tau \in t_{\{u,v\}}$. (25)

Constraints (1) and (2) ensure a one-to-one assignment between nodes of *V* and *V'*. Constraints (25) ensure that if two edges are assigned to each other and communication costs are in the forbidden set, then $y^{\tau}_{\{u,v\}\{u',v'\}} = 1$. This adds a penalty to the objective function to be minimized.

Example 7 illustrates the FSFA problem as a special case. The FSFA problem involves the assignment of discrete channels (or frequencies) to the transmitters of a radio network, such as a mobile telephone network. In that assignment, frequency separation is necessary to avoid interference by other transmitters to the signal received from the wanted transmitter at the reception points [68,69,70].

*Example 7.* The FSFA problem can be represented by a weighted undirected graph *G=(V, E)*. The problem looks for a mapping $f : V \to F$ where each $u \in V$ represents a transmitter of a radio network, and $F=\{0,1,...,|F|-1\}$ is a set of consecutive frequencies available for every vertex (transmitter) in *V*. The set of edges *E* represents those pairs of transmitters for which the assigned frequencies are constrained. The constraint arises because one of the transmitters can interfere with signals received from the other. For each $\{u,v\} \in E$ there is $t_{\{u,v\}} \in N^+$. The value of $t_{\{u,v\}}$ represents the highest separation between the frequencies assigned to the transmitter *u* and the one assigned to *v* that generates unacceptable interference. If we indicate with *f(u)* the frequency assigned to transmitter *u*, then if $|f(u)-f(v)|>t_{\{u,v\}}$, the interference involving the two transmitters is acceptable. The larger $|f(u)-f(v)|$ is the better the assignment, i.e., less interference. If however, $|f(u)-f(v)| \leq t_{\{u,v\}}$ there is a cost (penalty) equal to $p_{\{u,v\}} \in N^+$. The objective is to find an assignment *f* that minimizes the sum $p_{\{u,v\}}$ over all pairs $\{u,v\} \in E$ for which $|f(u)-f(v)| \leq t_{\{u,v\}}$. FSFA is a special case of the above formulation.

3. RELATIONSHIP WITH QUADRATIC ASSIGNMENT PROBLEM

The Quadratic assignment problem (QAP) is one of the most celebrated combinatorial problems with enormous applications [71]. There is a close relationship between the problems studied in this paper and QAP. Several variations of the QAP have been reported in the literature, (see [72] for a comprehensive survey). Those are the QAP minimizing total costs (QAP), the quadratic bottleneck assignment problem (QBAP), the quadratic semi-assignment problem (QSAP), the biquadratic assignment problem (BiQAP), the quadratic 3-dimensional assignment problem (Q3AP), and the multiobjective QAP (mQAP). The QAP and QSAP are special cases of the problem with output 2 and the objective function P2. The QBAP is a special case of the graph matching problem with output 2 and objective function P1. The BiQAP (also known as the *quartic assignment problem* [73,74]), and Q3AP are generalizations of QAP where the interaction is among several nodes of the graph, [72]. To save space, we will not go further into the details of these problems. However, we discuss QAP, and BQAP in the context of matching two graphs. In all cases proof is available. Due to saving space, we do not present the proof here, however, upon request the proof is available from the authors. QSAP is a straightforward variation of QAP.

The next section discusses situations in which there are interactions between several nodes of graphs. The proof is available upon request from the authors. Then in Part II, specific problems will be discussed. The classic book on complexity theory by Garey and Johnson [75] is a major source for definitions and understanding of these and other related problems that we use. For ease of presentation, if applicable, we refer to the *'problem index number'* addressed in [75].

4. GENERALIZATION TO MULTI-NODE-INTERACTION, AND MULTI-LAYER GRAPH MATCHING PROBLEMS

In the problems that we have discussed so far, interactions could occur only between two nodes of a graph. However, in many realistic situations' interactions among more than two nodes in the graph can occur, [73,74]. Furthermore, there are real problems, e.g., multi-layer crossing minimization problem [32], where the nodes of a graph are layered and thus the solution looks for several matching functions at the same time where the solution of each matching affects the whole configuration, [76]. Below we first consider the *interval graph completion (IGC) problem*, for which Even, Nao, Rao, and Schieber [28] presented an *indirect* IP formulation. It has applications in VLSI gate layout problems, [77]. A *natural* IP formulation has been proposed in [78]. However, we propose a simpler *natural* IP formulation.

*(GT35) Interval Graph Completion (IGC):*

Input: A connected undirected graph *G=(V, E).*

Output: A minimum cardinality set of edges *F* such that $G=(V,E_s)$ is an interval graph and $E_s = E \cup F$.

A graph is an interval graph if and only if there is a linear ordering of the vertices such that if vertex *u* with index *u'* has an edge to vertex *v* with index *v'*, where *u'<v'*, then every vertex whose index is between *u'* and *v'* also has an edge to vertex *v*, [28]. Define a complete undirected graph *G=(V',E')* where *|V|=|V'|* with nodes numbered *1,2,...,|V|*. The IGC problem is to find a one-to-one assignment, $f:V \to V'$ where the

condition of an interval graph is satisfied. Below we give an IP formulation of the problem. Here, $y_e = 1$ if an edge $e \in \overline{E}$ is not among minimum cardinality set $F$, and zero otherwise.

Min $\sum_{e \in \overline{E}} y_e$,

s.t. $\sum_{u' \in V'} x_{uu'} = 1$, for $u \in V$,   $\sum_{u \in V} x_{uu'} = 1$, for $u' \in V'$,   (26)

$x_{uu'} + x_{w'} + x_{zz'} \leq y_{\{z,v\}} + 2$, for $\{u,v\} \in E$, and $u' < z' < v'$ and $\{z,v\} \in \overline{E}$,   (27)

Constraints (26) ensure a one-to-one assignment between $V$ and $V'$. Constraints (27) ensure that for an edge $\{u,v\}$ where $u$ is assigned to $u'$ and $v$ is assigned to $v'$ with $u' < v'$ if a node $z$ is assigned to a node $z'$ with $u' < z' < v'$ and if there is no an edge $\{z,v\}$ in $E$ then the value of $y_{\{z,v\}}$ must be equal to 1 and thus edge $\{z,v\}$ is added to the set $F$. The objective function minimizes the cardinality of the set $F$. The IGC is equivalent to the *profile minimization problem (PMP)*, [79]. The PMP is discussed in Part II.

Now, we discuss the problem of *multi-layer crossing minimization* that has many applications including combinatorial geometry, the theory of VSLI, graph drawing, and computational biology [25,80,81,82,83,84]. Nollenberg and Wolff [85] have recently presented an IP formulation for a special case of this problem.

Given a directed graph $G=(V, A)$ the set of nodes $V$ is partitioned into $p$ disjoint sets, where $V_l \cap V_k \neq \varnothing$, for $l,k=1,...,p$, $l \neq k$ and $V = \bigcup_{l=1}^{p} V_l$ with $|V_l| = n_l$. Define an arc $(u,v) \in A$ for $u \in V_l$ and $v \in V_{l+1}$, for some $l=1,...,p-1$ (where each set of nodes $V_l$ is placed on a straight line, i.e., layer $l$). We look for a permutation of vertices $V_l$ in each layer $l$ such that the number of crossing arcs is minimized. Here, the set of arcs is partitioned into $A_1,...,A_{p-1}$ where $A_l$ is the set of arcs between nodes $V_l$ and $V_{l+1}$ for $l=1,...,p-1$.

*Multi-layer Crossing Minimization (MLCM):*

Input: A multi-layered graph $G=(V, A)$ where the set of nodes $V$ is partitioned into $p$ disjoint sets, $V_1,...,V_p$, for $l=1,...,p$.

Output: A permutation of vertices $V_l$ in each layer $l$ such that the number of crossing arcs is minimized.

If $p=2$ the problem is known as 2-layer crossing minimization, which is NP-hard [86]. Several IP formulations of the 2-layer problem have been provided, [37,87]. The MLCM problem is NP-hard even if the permutation of one layer is fixed, known as *one-sided* crossing minimization [88]. Recently, several other special cases of the MLCM problem have shown to be NP-hard, [89,90]. In most heuristics and/or exact algorithms, the 2-layer version of the problem is extensively used with or without permutation of one side fixed, [32]. Carpano, [91] stated that the difficulty with dealing with the general MLCM involves not only finding the best permutation for each layer concerning its adjacent layers but also the effect of this choice on the configuration as a whole. In that spirit, an IP formulation was presented in [32].

Define a one-to-one function $f_l : V_l \rightarrow \{1,...,n_l\}$ for each layer $l$. Let $x_{ui}^l = 1$ if node $u \in V_l$ is assigned to $i \in \{1,...,n_l\}$, and 0 otherwise. Two arcs $(u,u'),(v,v') \in A_l$ (for $u,v \in V_l, u',v' \in V_{l+1}$) cross iff either $f_l(u) < f_l(v)$ and $f_{l+1}(v') < f_{l+1}(u')$ or $f_l(v) < f_l(u)$ and $f_{l+1}(u') < f_{l+1}(v')$. Let $y_{(u,u')(v,v')}^l = 1$ iff two arcs $(u,u')$ and $(v,v')$ in layer $l$ cross, and 0 otherwise. Our new IP formulation of the problem is as follows.

Min $\sum_{l=1}^{p-1} \sum_{(u,u'),(v,v') \in A_l} y_{(u,u')(v,v')}^l$

s.t. $\sum_{i=1}^{n_l} x_{ui}^l = 1$, for $l \in \{1,...,p\}, u \in V_l$,   $\sum_{u \in V_l} x_{ui}^l = 1$, for $l \in \{1,...,p\}, i \in \{1,...,n_l\}$,   (28)

$x_{ui}^l + x_{vj}^l + x_{v'i'}^{l+1} + x_{u'j'}^{l+1} \leq y_{(u,u')(v,v')}^l + 3$, for $l \in \{1,...,p-1\}, (u,u'),(v,v') \in A_l$,

for $i, j \in \{1,...,n_l\}$, and $i', j' \in \{1,...,n_{l+1}\}$, and $i < j, i' < j'$,   (29)

$$x^l_{uj} + x^l_{vi} + x^{l+1}_{u'i'} + x^{l+1}_{v'j'} \leq y^l_{(u,u')(v,v')} + 3, \quad \text{for } l \in \{1,\ldots, p-1\}, (u,u'),(v,v') \in A_l,$$
$$\text{for } i, j \in \{1,\ldots, n_l\}, \text{ and } i', j' \in \{1,\ldots, n_{l+1}\}, \text{ and } i < j, i' < j'. \tag{30}$$

Constraints (28) represent a one-to-one assignment between a set of layered nodes $V_l$ and $\{1,\ldots, n_l\}$, $l=1,\ldots,p$. Constraints (29-30) ensure a penalty value of 1 is added to the objective function if two edges cross. The objective function therefore minimizes the sum of all crossings.

## PART II

Graph theoretic approaches for solving difficult problems have been an important practical and theoretical research area for several decades. Graphs have been applied to almost every discipline of sciences, engineering, and social sciences. Some of the major applications in graph matching are pattern analysis, social network analysis, homeland security, telecommunications, and manufacturing [10, 27,29,49,53,92,93,94,95,96,97,98,99,100,101,102]. In Part II we present IP formulations for a wide range of difficult problems, problems 1-12 from the list we presented earlier, using the framework presented in Part I. For some of these problems, we have not seen any *natural* IP formulations in the literature and for some others, our formulations provide alternative approaches.

5. Integer Programming Formulations of Graph Problems

This section presents *natural* integer programming formulation for problems 1-12 in the list that we presented earlier (refer to the survey papers [2,5,92] for comprehensive discussion and applications). Also, the classic book on complexity theory by Garey and Johnson [75] is a major source for definitions and understanding of these and other related graph problems. We may discuss the *recognition* (*feasibility*) and/or the *optimization* problems separately based on the appropriateness and availability of limited space. For simplicity, the formulas in Table 2 are used over and over when we present the formulations. To save space we do not present proof here, however, upon request the proofs are available from the authors.

We start the discussion with some variations of the TSP. For a comprehensive survey of variants of TSP refer to [103]. We discuss a variation, known as *k*-TSP, that includes TSP and the Bottleneck TSP as special cases, [42,43]. We discuss the *k*-TSP for directed graphs. Formulations can be easily modified for undirected graphs. Our formulation here is different and more general than those in the literature, (see [7,46] for classification of TSP formulations). Also note that in Part I, we gave several new IP formulations of *k-TSP* and *induced k-TSP* for undirected graphs. Similar formulations can be given for directed graphs, however, here in Part II we present different formulas for these problems. To save space our concentration will be on *k-TSP* for directed graphs. With minor modifications, the formulation can be extended to *induced k-TSP*.

-------------------------------
Table 2 about here
-------------------------------

5.1. Variations of TSP

*(ND22, ND23, ND24) Some Variations of TSP:*

Input: Directed graph $G=(V,A)$, distance $w_{(u,v)}$ for arc $(u,v) \in A$, and positive integers $k \leq n$, and $K$.

Output A: Is there a circuit of *G* with *k* nodes?
Output B: Is there a circuit of *G* with *k* nodes with a total travel distance of *K* or less?
Output C: Is there a circuit of *G* with *k* nodes with the longest distance between any two cities less than or equal to *K*?

We first define a *circle graph G'=(V', A')* with *k* nodes similar to Figure 1. In that, for $u'=1,\ldots,k-1$ we have $(k,1),(u',u'+1) \in A'$. The problem is for an assignment $f: V \to V'$, where each element of *V* may be assigned to an element of *V'*, and each element of *V'* must be assigned to an element of *V*. This is a *subgraph isomorphism problem.* If *k=n,* we have the Hamiltonian circuit, i.e., the TSP, and the Bottleneck TSP, respectively, (see [104,105] for recent optimal procedures of TSP and Bottleneck TSP). The formulation of the problem for Output A can be stated as follows:

(a2),(b1), and $x_{uu'} \leq \sum_{v \in N(u)} x_{v(u'+1)}$, for $u \in V, u' \in V \setminus \{k\}$, and $x_{uk} \leq \sum_{v \in N(u)} x_{v1}$, for $u \in V$.

Constraints (a2) and (b1), taken from Table 2, ensure that a node in $V$ may be assigned to a node in $V'$ and a node in $V'$ must be assigned to exactly one node in $V$. The first set of inequalities ensures that if a node $u$ is assigned to any node $u'$, except node $k$, in $V'$ then a node $v$ where $(u,v)$ is an arc in $G$ must be assigned to node $u'+1$. The last set of inequalities endures that for a node $u$ in $V$ assigned to node $k$ in $V'$, there must be a node $v$ in $V$ such that $(u,v)$ is an arc in $G$ and $v$ is assigned to node 1. If there is a solution to this formulation then we have found a circuit of $k$ nodes in graph $G$, otherwise no such circuit exists in graph $G$. Formulation of the problem for Output B can be stated as follows.

(a2),(b1),(g) and $x_{uu'} \leq \sum_{v \in N(u)} x_{v(u'+1)}$, for $u \in V, u' \in V \setminus \{k\}$, and $x_{uk} \leq \sum_{v \in N(u)} x_{v1}$, for $u \in V$.

$$\sum_{u \in V} \sum_{v \in N(u)} \sum_{u' \in V \setminus \{k\}} \sum_{u'+1} w_{(u,v)} y_{(u,v)(u',u'+1)} + \sum_{u \in V} \sum_{v \in N(u)} w_{(u,v)} y_{(u,v)(k,1)} \leq K.$$

Here to the set of feasible solutions in output A, constraints (g), from Table 2, are added. Constraints (g) ensure that if two arcs $(u,v)$ and $(u',v')$ are assigned to each other then $y_{(u,v)(u',v')} = 1$. The LHS of the last inequality adds up all distances in creating a circuit that needs to be within a bound $K$. Note that the LHS is in the form of P2. Thus, if *Min K* is added to the formulation, we find the smallest $K$ that satisfies Output B. Formulation of the problem for Output C can be given as a bottleneck optimization. A feasible solution is the same as for output B, however, the last inequality is replaced by,

$w_{(u,v)} y_{(u,v)(u',v')} \leq K$, for $(u,v) \in A$, and $(u',v') \in A'$.

Note that here also the LHS is in the form of P1. Thus, if *Min K* is added to the formulation, we find the smallest $K$ that satisfies output C. Next, we present several linear arrangement problems. Bandwidth optimization on a graph which has a variety of applications, [26,106,107,108], is discussed first. IP formulations have been given in [33,35], however, our formulation differs from theirs.

5.2. Linear Arrangement Problems on Graphs

*(GT40) Bandwidth Problem (BP):*

Input: Undirected graph $G=(V,E)$, and a positive integer $K \leq n$.
Output: Is there a linear ordering of $V$ with the bandwidth $K$ or less, i.e., a one-to-one function $f: V \to \{1, \ldots, n\}$ such that, all edges $\{u,v\} \in E$, $|f(u) - f(v)| \leq K$?

Define an undirected graph $G'=(V', E')$ with $|V'|=n$, where the nodes are numbered $1, \ldots, n$. For nodes $u', v' \in V'$ define an edge iff $|u'-v'| \leq K$. Now, we have $E' = \{\{u',v'\}: u',v' \in V', \text{ and } w_{\{u',v'\}} = |u'-v'| \leq K\}$.

The BP feasibility problem can be stated as a problem of finding a one-to-one assignment, $f: V \to V'$, such that for each $\{u,v\} \in E$ we have $\{f(u), f(v)\} \in E'$. An IP formulation derives from satisfying constraints (a1),(a2), and (c1) in Table 2. In the optimization version of the BP, we look for a one-to-one assignment $f: V \to \{1, \ldots, n\}$ to minimize $K$. In that, $G'$ is a complete graph and the objective function is $Min_f \{Max_{\{u,v\} \in E} |f(u) - f(v)|\}$ which is in the form of P1. IP formulation is as follows:

*Min K*, s.t. (a1),(a2),(f), and $w_{\{u',v'\}} y_{\{u,v\}\{u',v'\}} \leq K$, for $\{u,v\} \in E$, and $u', v' \in V'$.

Inequalities (f) assign a value of 1 to $y_{\{u,v\}\{u',v'\}}$ when an edge $\{u,v\} \in E$ is assigned to an edge $\{u',v'\} \in E'$. The last set of inequalities ensures that a value of $|f(u)-f(v)|$ is assigned to edge $\{u,v\}$.

*(GT41) Directed Bandwidth Problem (DBP):*

Input: Directed graph $G=(V,A)$, and a positive integer $K \leq n$.
Output: Is there a one-to-one function $f: V \to \{1, \ldots, n\}$ such that, for all arcs $(u,v) \in A$, $f(u) < f(v)$ and $f(v) - f(u) \leq K$?

Similar to the case of BP define a digraph $G'=(V', A')$. Define an arc between two nodes $u',v' \in V'$ if $v'>u'$ and $v'-u' \leq K$. Now, we have $A'=\{(u',v'): u',v' \in V',$ with $w_{(u',v')} = v'-u' \leq K\}$. The DBP feasibility problem can be stated as a problem of finding a one-to-one assignment, $f: V \to V'$, such that for each $(u,v) \in A$ we have $(f(u), f(v)) \in A'$. IP formulation of the DBP derives from satisfying constraints (a1),(a2),(d1), and (e) in Table 2. In the optimization version of the DBP, graph $G'$ is a complete graph. We are looking for a one-to-one assignment $f: V \to \{1,\ldots,n\}$ to minimize $K$. The objective function is $Min_f \{Max_{(u,v) \in A}\{f(v)-f(u)\}\}$ which is in the form of P1. IP formulation is as follows:

  $Min\ K$, s.t. (a1), (a2), (e), (g), and $w_{(u',v')} y_{(u,v)(u',v')} \leq K$, for $(u,v) \in A$, and $(u',v') \in A'$.

Next, we discuss the linear arrangement problem (LAP) which has a variety of applications, [2,109,110].

*(GT42) Linear Arrangement Problem (LAP):*

Input: Undirected graph $G=(V,E)$ with weight $w_e$ for each $e \in E$, and a positive integer $K$.

Output: Is there a one-to-one function $f: V \to \{1,\ldots,n\}$ such that $\sum_{\{u,v\} \in E} w_{\{u,v\}} |f(u)-f(v)| \leq K$?

Define an undirected graph $G'=(V',E')$ with $|V'|=n$, and nodes numbered $1,\ldots,n$. For two nodes $u',v' \in V'$ let $w_{\{u',v'\}} = |u'-v'|$. Find a one-to-one assignment $f: V \to V'$ such that $\sum_{\{u,v\} \in E} w_{\{u,v\}} |f(u)-f(v)| \leq K$. An IP formulation can be stated as follows which is different from those in [36,37].

  (a1), (a2), (f), and $\sum_{\{u,v\} \in E, \{u',v'\} \in E'} (w_{\{u,v\}} w_{\{u',v'\}} y_{\{u,v\}\{u',v'\}}) \leq K$.

The optimization of the problem is in the form of P2. We need to add *Min K* to the feasibility problem.

*(GT43) Directed Linear Arrangement Problem (DLAP):*

Input: Directed graph $G=(V,A)$ with weight $w_a$ for each $a \in A$, and a positive integer $K$.

Output: Is there a one-to-one function $f: V \to \{1,\ldots,n\}$ such that $f(u)<f(v)$ whenever $(u,v) \in A$ and such that $\sum_{(u,v) \in A} w_{(u,v)}(f(v)-f(u)) \leq K$?

Define a complete digraph $G'=(V',A')$ where $|V'|=n$ with nodes numbered $1,\ldots,n$. For $u',v' \in V'$ define an arc $(u',v')$ if $v'>u'$, and let $w_a = v'-u'$. We want to find a one-to-one assignment $f: V \to V'$ such that for each arc $(u,v) \in A$ if $u$ is assigned to the node $u' \in V'$ and $v$ is assigned to the node $v' \in V'$ then there is an arc $(u',v') \in A'$ such that $\sum_{(u,v) \in A} w_{(u,v)} (f(v)-f(u)) \leq K$. An IP formulation of the problem is stated as follows.

  (a1),(a2),(e),(g) and $\sum_{(u,v) \in A, (u',v') \in A'} (w_{(u,v)} w_{(u',v')} y_{(u,v)(u',v')}) \leq K$.

In the optimization version of the problem, we look for an assignment that minimizes the total cost. Thus, we only need to add *Min K* to the feasibility problem that is in the form of P2. Next, we discuss the *profile minimization problem* which has many applications, [79,111,112]. It is closely related to BP and LAP. It has been shown in [79] that the PMP is equivalent to the IGC problem (IGC was discussed in Part I).

*Profile Minimization Problem (PMP):*

Input: Undirected graph $G=(V, E)$.

Output: A one-to-one function $f: V \to \{1,\ldots,|V|\}$ such that $K = \sum_{u \in V} Max_{v \in N[u]}\{f(u)-f(v)\}$ is minimized.

Define a complete undirected graph $G'=(V',E')$ with $|V'|=\{1,\ldots,|V|\}$. We want to find a one-to-one assignment $f: V \to V'$ that minimizes the value of $K$. There is a clear similarity between PMP, BP, and LAP. PMP has the elements of both BP and LAP, (see [7]). For $u \in V$ let $c(u, f(u)) = f(u)$, and for $\{u,v\} \in E$ let $d(\{u,v\},\{f(u),f(v)\}) = -Min[f(u),f(v)]$. The objective of PMP is

$Min_f \{\sum_{u \in V} Max_{v \in N(u)} \{c(u,f(u)) + d(\{u,v\},\{f(u),f(v)\})\}\}$.

An IP formulation of PMP can be stated as follows, which is in the form of P3:

$Min_f \sum_{u \in V} K_u$, s.t. (a1),(a2),(f), and

$\sum_{u' \in V'} c(u,u') x_{uu'} + \sum_{u' \in V'} \sum_{v' \in V'} d(\{u,v\},\{u',v'\}) y_{\{u,v\}\{u',v'\}} \leq K_u$, for $u \in V, v \in N(u)$.

We next consider the *min-cut linear arrangement problem*, (see [6] for recent discussion and complexity of special cases). It has applications in telecommunication [65], computer-aided design, automatic graph drawing, protein engineering, optimization of rail systems [2], and computational logic [66].

*(GT44) Minimum Cut Linear Arrangement Problem (MCLAP):*

Input: Undirected graph *G=(V, E)*.

Output: Find the smallest value of *K* and a one-to-one function $f: V \to \{1,\ldots,n\}$ such that for all $1 \leq i \leq n$,

$|\{\{u,v\} \in E, f(u) \leq i < f(v)\}| \leq K$.

Define a complete undirected graph *G'=(V', E')* with *|V'|=n* and let the nodes be numbered *1,…,n*. In the MCLAP feasibility problem, we want to find a one-to-one assignment between nodes graphs *G* and *G'*, such that given any node $i' \in V'$, the cardinality of all edges *{u,v}* in *E* with one end *u* assigned to node $u' \in V'$ ($u' \leq i'$) and the other end *v* assigned to the node $v' \in V'$ ($v' > i'$) is within the limit *K*. An IP formulation of the feasibility problem can be given as follows:

(a1),(a2),(f), and $\sum_{\{u,v\} \in E} \sum_{u' \leq i', v' > i'} y_{\{u,v\}\{u',v'\}} \leq K$, for $i' \in V'$

The formulation can be extended to the optimization of MCLAP by adding,

$Min_f \{Max_{i' \in V'} \{\sum_{\{u,v\} \in E} \sum_{u' \leq i', v' > i'} y_{\{u,v\}\{u',v'\}}\}\}$.

An IP formulation is derived by adding *Min K* to the feasibility problem. The objective function is in the form of P7. In that, we have *c(u,f(u))=0* for all *u* in *V*, and *d({u,v},{f(u),f(v)})=1* for all *{u,v}* in *E*. In the next section, we discuss graph coloring variations. We first consider the *graph* and *subgraph isomorphism problems*. The complexity of the graph isomorphism problem is still open [11], although the subgraph isomorphism problem is known to be NP-hard [75].

5.3. Graph Coloring Variations

*(A13, OPEN 1, page 285) Graph Isomorphism (GI):*

Input: Undirected graphs *G=(V,E)* and *G'=(V',E')*, with *|V|=|V'|=n*.

Output: Are *G* and *G'* isomorphic, i.e., is there a one-to-one onto function $f: V \to V'$ such that $\{u,v\} \in E$ if and only if $\{f(u), f(v)\} \in E'$?

An IP formulation of the GI derives from satisfying (a1),(a2),(c1), and (c2) in Table 2. Next, we discuss the *subgraph isomorphism problem (SI)* and the *induced subgraph isomorphism problem (ISI)*. A subgraph of a graph *G* is a graph whose vertex set is a subset of that of *G*, and whose edges are a subset of that of *G* restricted to this subset. A subgraph *H* of a graph *G* is said to be *induced* (or full) if for any pair of vertices *u* and *v* of *H*, *{u,v}* is an edge of *H* if and only if *{u,v}* is an edge of *G*. In other words, *H* is an induced subgraph of *G* if it has exactly the edges that appear in *G* over the same vertex set (refer to example 1 for pictorial illustration).

*(GT48) Subgraph Isomorphism (SI):*

Input: Undirected graphs *G=(V,E)* and *G'=(V',E')*, with $|V| = n \geq |V'| = n'$.

Output: Does *G* contain a subgraph isomorphic to *G'*, i.e., a subset $S \subseteq V$ and a subset $L \subseteq E$ such that *|S|=|V'|*, *|L|=|E'|* and there exists a one-to-one function $f: V' \to S$ satisfying $\{u',v'\} \in E'$ if and only if $\{f(u'), f(v')\} \in L$?

*Induced Subgraph Isomorphism (ISI):*

Input: Undirected graphs *G=(V,E)* and *G'=(V',E')*, with $|V| = n \geq |V'| = n'$.

Output: Does $G$ contain a subgraph isomorphic to $G'$, i.e., a subset $S \subseteq V$ and a subset $L \subseteq E$ such that $|S|=|V'|$, $|L|=|E'|$ and there exists a one-to-one function $f: V' \to S$ satisfying $\{u',v'\} \in E'$ if and only if $\{f(u'), f(v')\} \in E$ ?

Note that as was discussed in example 1, Part I, in SI we do not rule out the possibility of existing $\{u,v\} \in E \setminus L$ such that $f(u')=u$ and $f(v')=v$ for some $\{u',v'\} \in E'$. An IP formulation of SI derives from satisfying (a2),(b1), and (c2) in Table 2, and an IP formulation of ISI derives from satisfying (a2),(b1),(c1), and (c2). All discussions on GI and SI with the undirected graphs can be extended to the directed graphs straightforwardly. Thus, to save space we will not present them here. Below we discuss the largest common subgraph problem (LCS) that has a variety of applications, especially in data analysis, [113,114,115]. An IP formulation was presented in [116]. We present a simpler formulation with significantly fewer variables.

*(GT49) Largest Common Subgraph Problem (LCS):*
Input: Undirected graphs $G=(V,E)$ and $G'=(V',E')$.
Output: Find the largest subsets $L \subseteq E$ and $L' \subseteq E'$ with $|L|=|L'|$ such that the subgraphs $G_1 = (V, L)$ and $G_1' = (V', L')$ are isomorphic.

Consider an assignment $f: V \to V'$ where a node $u \in V$ can be assigned to only one node $u' \in V'$ and vice versa. Here, two edges, $\{u,v\} \in E$ and $\{u',v'\} \in E'$ are assigned to each other only if we have either $x_{uu'} + x_{vv'} = 1$ or $x_{uv'} + x_{vu'} = 1$, but not both. Let $w_e = w_{e'} = 1$ for all $e \in E$ and $e' \in E'$. For two edges $\{u,v\} \in E$ and $\{u',v'\} \in E'$, we have $w_{\{u,v\}} w_{\{u',v'\}} (x_{uu'} x_{vv'} + x_{uv'} x_{vu'}) = 1$ iff they are assigned to each other. A quadratic IP formulation of the problem can be stated as follows:

$Max \sum_{u \in V} \sum_{u' \in V'} \sum_{v \in V} \sum_{v' \in V'} w_{\{u,v\}} w_{\{u',v'\}} (x_{uu'} x_{vv'} + x_{uv'} x_{vu'})$, s.t. (b1),(b2).

To transfer the quadratic formulation to a linear integer program, we let $y_{\{u,v\}\{u',v'\}} = 1$ if an edge $\{u,v\} \in E$ is assigned to an edge $\{u',v'\} \in E'$ with $u$ assigned to $u'$ and $v$ assigned to $v'$, and 0 otherwise. We let $z_{\{u,v\}\{u',v'\}} = 1$ if an edge $\{u,v\} \in E$ is assigned to the edge $\{u',v'\} \in E'$ with $u$ assigned to $v'$ and $v$ assigned to $u'$, and 0 otherwise. An IP formulation can be stated as follows where objective function is in the form of P2.

$Min \sum_{\{u,v\} \in E, \{u',v'\} \in E'} (y_{\{u,v\}\{u',v'\}} + z_{\{u,v\}\{u',v'\}})$ , s.t. (b1),(b2), and

$y_{\{u,v\}\{u',v'\}} \leq x_{uu'}, y_{\{u,v\}\{u',v'\}} \leq x_{vv'}, z_{\{u,v\}\{u',v'\}} \leq x_{uv'}, z_{\{u,v\}\{u',v'\}} \leq x_{vu'}$, for $\{u,v\} \in E, \{u',v'\} \in E'$.

Next, we present maximum subgraph matching (MSM) with a variety of applications, [117,118,119].

*(GT50) Maximum Subgraph Matching (MSM):*
Input: Directed graphs $G=(V,A)$ and $G'=(V',A')$.
Output: Find the largest subsets $R \subseteq V \times V'$ such that for all $<u,u'>,<v,v'> \in R$, $(u,v) \in A$ if and only if $(u',v') \in A'$.

Here, we have many-to-many matching. A slightly different version of the problem is that the two generated subgraphs must preserve all arc relations and thus must be *isomorphic*. Authors in [120] originally introduced the MSM problem with no references to isomorphism, stating, *"find the maximal matches, where a match is a correspondence (many-many relation) between a subgraph H of G and a subgraph H' of G', which preserves the relation."* In the following, first, we discuss the *isomorphic* version of the problem (MISM) then we discuss MSM as stated in GT50. Consider an assignment $f: V \to V'$ with the assumption that the node $u \in V$ can be assigned to only one node $u' \in V'$, and vice versa. We look for the largest subsets $S \subseteq V$ and $S' \subseteq V'$ with $|S|=|S'|$ to be assigned to each other and such that for all $u,v \in S$ and $u',v' \in S'$, $(u,v) \in A$ if and only if $(u',v') \in A'$. The two subsets of nodes $S \subseteq V$ and $S' \subseteq V'$, respectively, create *induced* subgraphs in $G$ and $G'$.

*Maximum Induced Subgraph Matching (MISM):*
Input: Directed graphs $G=(V,A)$ and $G'=(V',A')$.

Output: Find the largest subsets $S \subseteq V$ and $S' \subseteq V'$ with $|S|=|S'|$ and a one-to-one function $f : S \to S'$ such that for all $u,v \in S$ and $u',v' \in S'$, $(u,v) \in A$ if and only if $(u',v') \in A'$.

An IP formulation is as follows.

$Max \sum_{u \in V} \sum_{u' \in V'} x_{uu'}$, s.t. (b1),(b2),(d1),(d2),(e).

Now, an IP formulation for the MSM problem, GT50, can be stated as follows.

$Max \sum_{u \in V} \sum_{u' \in V'} x_{uu'}$, s.t. (d1),(d2),(e),(h2),(i2).

The objective functions in the above problems are in the form of P2. In the next several subsections, we consider the graph coloring and several related problems and their generalizations. The relationship among these problems is discussed, and IP formulations are introduced. The problems have enormous applications (see for example, [20,56,121,122,123,124,125,126]).

*(GT4) Graph K-Colorability (GKC):*

Input: Undirected graph $G=(V,E)$, and an integer $K \leq n$.
Output: Does there exist a function $f : V \to \{1,\ldots,K\}$ such that $f(u) \neq f(v)$ whenever $\{u,v\} \in E$?

*Graph Coloring (GC):*

Input: Undirected graph $G=(V, E)$.
Output: Find the smallest value of $K$, a function $f : V \to \{1,\ldots,K\}$ such that $f(u) \neq f(v)$ whenever $\{u,v\} \in E$.

A generalization of the GKC is a graph *homomorphism* stated as follows, GT52. An extension to the directed graph is also presented.

*(GT52) Graph Homomorphism (GH):*

Input: Undirected graphs $G=(V,E)$ and $G'=(V',E')$.
Output: Find a *homomorphism* from $G$ to $G'$, i.e., find a function $f : V \to V'$ such that $\{f(u), f(v)\} \in E'$ whenever $\{u,v\} \in E$.

*Directed Graph Homomorphism (DGH):*

Input: Directed graphs $G=(V,A)$ and $G'=(V',A')$.
Output: Find a *homomorphism* from $G$ to $G'$, i.e., find a function $f : V \to V'$ such that $(f(u), f(v)) \in A'$ whenever $(u,v) \in A$.

An IP formulation of the GH problem derives from satisfying (a1),(c1), and (i1) in Table 2, and for DGH it derives from satisfying (a1),(d1),(e) and (i2). Consider now the GKC, and define a complete undirected graph $G'=(V', E')$ where $|V'|=K$ with the nodes numbered $1,\ldots, K$. Now, GKC seeks an assignment, $f : V \to V'$, such that $\{f(u), f(v)\} \in E'$ whenever $\{u,v\} \in E$. Since $G'$ is a complete graph, this is the same as finding an assignment, $f : V \to V'$ such that the endpoints of an edge $\{u,v\} \in E$ cannot be assigned simultaneously to one node in $G'$. An IP formulation of GKC is derived from satisfying (a1) and (i1) in Table 2. In the *graph coloring optimization problem (GC)*, the goal is to find the smallest number of colors to assign to elements of the set *V*. Given an upper bound for the number of nodes $V'$ (e.g., $|V'|=n$), similar to GKC define a complete graph $G'$. The IP formulation for the GC problem is now stated as follows:

*Min K*, s.t. (a1),(i1), and $u' x_{uu'} \leq K$, for $u \in V, u' \in V'$.

Consider now a generalization of the coloring problem that we call the *maximum (minimum) weight subset coloring problem* (MWSCP). The problem both in maximum and minimum format has many applications. In MWSCP, if node $u \in V$ receives a color $u' \in V'$ (where *V'* is the set of colors), then there is a benefit (or a cost) of *c(u,u')* (refer to [56] for approximation algorithms for special cases of these problems). For a given subset $S \subseteq V$ and a function $f : S \to \{1,\ldots,K\}$, define $Z = \sum_{u \in S} c(u, f(u))$ to be the weight of *S*. Now, MWSCP can be stated as follows:

*Maximum (Minimum) Weight Subset Coloring Problem (MWSCP):*

Input: Undirected graph $G=(V,E)$, and an integer $K \leq n$.

Output: Find a maximum (or minimum) weight subset $S \subseteq V$, and the smallest positive integer $l \leq K$, and a function $f: S \rightarrow \{1,\ldots,l\}$ such that $f(u) \neq f(v)$ whenever $\{u,v\} \in E$ and $u,v \in S$.

In MWSCP, we have a bi-objective optimization problem. We discussed the problem for the maximum weight subset. It can be easily modified for the minimum case. In MWSCP, we look for a maximum weight subset $S \subseteq V$ that can be colored by $K$ colors. Let $Z^*$ be the maximum weight. We then want to find the smallest number of colors, $l$, that admits $Z^*$. Define a complete graph $G'=(V',E')$ with $V'=\{1,...,K\}$. Now, IP formulations can be given as follows. The IP (A) is in the form of P2, and (B) is in the form of P1.

(A) $Max\ Z = \sum_{u \in V} \sum_{u' \in V'} c(u,u') x_{uu'}$, s.t. (b1),(i1).

To find the smallest number of colors, i.e., the *strength* of $G$, we solve the following IP.

(B) $Min\ l$, s.t. (b1),(i1), and $u' x_{uu'} \leq l$, for $u \in V, u' \in V'$, $l \geq 0$, and

$$\sum_{u \in V} \sum_{u' \in V'} c(u,u') x_{uu'} = Z^*.$$

Several important combinatorial optimization problems are special cases of MWSCP:
(1) If $K=1$ then we have a single color and thus $G'$ has a single vertex with no edge. For each node $u \in V$ let $w_u$ be a given weight and define $c(u, f(u)) = w_u$. Now, IP (A) solves the well-known *maximum weight independent set* problem.
(2) If $K=n$, then we have $S=V$. Given an assignment $f: V \rightarrow \{1,\ldots,K\}$ that satisfies the condition of MWSCP, for node $u \in V$ let $c(u, f(u)) = f(u)$. Here, the weight of $c(u,f(u))$ is exactly equal to the integer number of nodes assigned to $u$. Now if we minimize IP (A) instead of maximizing, then we have solved the *optimum cost chromatic partition (OCCP)* [127,128,129], which has the *sum coloring (SC)* [130] as a special case. Both OCCP and SC problems have numerous applications in scheduling theory, resource allocation, and VSLI design [98]. IP (B) then solves for the *strength* of graph $G$. Note that these problems can be considered as special cases of the *Graph Grundy Number* (GGN, GT56, [75]), which is defined on digraphs. Our formulation can easily be modified for GGN (see [93] for the survey of results).

Another generalization of the graph coloring is as follows, known as *weighted vertex coloring (WVCP)*. Its special cases have been discussed in the literature and have applications in batch scheduling, transmission of real-time messages in metropolitan networks, and dynamic storage allocation [54,55]. IP formulation was presented in [57].

*Weighted Vertex Coloring Problem (WVCP):*

Input: Undirected graph $G=(V,E)$, and positive weight $w_u$ for each $u \in V$.

Output: Find a function $f: V \rightarrow \{1,\ldots,K\}$ for some $K \leq n$ such that $f(u) \neq f(v)$ whenever $\{u,v\} \in E$ and $u,v \in V$, to minimize $Z = \sum_{1 \leq u' \leq K} Max_{u \in V, f(u)=u'}\{w_u\}$.

Let $K_{max}$ be an upper bound to the optimal value of $K$ (e.g., $K_{max} = n$), now define a complete graph $G'=(V',E')$ where $V' = \{1,\ldots,K_{max}\}$. The problem is to find a function $f: V \circledR V'$ that minimizes $Z$. Here, the objective function is a special case of P5. An IP formulation of WVCP is presented below. In the special case when all weights are equal to 1, we have the GC problem. However, the IP formulation of the special case of GC is different from the one we presented earlier (see [131] for a similar formulation of a special case). In this formulation we have $c(u, f(u)) = w_u$.

$Min\ \sum_{u' \in V'} K_{u'}$, s.t. (a1),(i1), and $c(u,u') x_{uu'} \leq K_{u'}$, for $u \in V, u' \in V'$.

An important graph optimization problem with enormous applications is the *maximum clique partition problem* (MCPP). It has the *graph partitioning problem* (GPP) as a special case. Applications of MCPP and GPP include clustering and pattern analysis (see [21,22,23,101,102,132,133,134,135,136] for recent applications). The problem is stated as follows:

*Maximum Clique Partition Problem (MCPP):*

Input: Undirected graph $G=(V,E)$, weight $w_e$ (a real number) for each $e \in E$.

Output: Find a partition, $V = \{V_1, \ldots, V_k\}$, of vertices into disjoint cliques to maximize
$Z = \sum_{1 \leq l \leq K} \sum_{u,v \in V_l, \{u,v\} \in E} w_{\{u,v\}}$.

The value of $K$ that admits the maximum in the problem is called the *maximum clique partition number*, $\bar{\chi}_{max}(G)$. Authors in [137,138] presented an IP formulation of the problem. Our formulation is based on the GC concept and is different from theirs. Let $\bar{G} = (V, \bar{E})$ be the complement of graph $G=(V, E)$. Now, the MCPP is equivalent to the problem of partitioning vertices into *independent* sets (or coloring of vertices) where the sum over all weight of edges, $e \notin \bar{E}$, i.e., between pair of vertices in each independent set is maximized. Let $K_{max}$ be an upper bound to $\bar{\chi}_{max}(G)$ (e.g., $K_{max} = n$). Define a graph $G'=(V', E')$ where $V' = \{1, \ldots, K_{max}\}$ and $E'$ contains only self-loop edges for each node of $V'$. We look for a function $f: V \rightarrow V'$ such that $f(u) \neq f(v)$ whenever $\{u,v\} \in \bar{E}$ and such that $Z$ is maximized. A set of nodes of $G$ that are assigned to a node $u'$ in $G'$ create an independent set in $\bar{G}$, and thus a clique in $G$. An IP formulation can be given as follows, which is in the form of P2.

*Max* $\sum_{u' \in V'} \sum_{u,v \in V, \{u,v\} \in E} w_{\{u,v\}} y_{\{u,v\}\{u',u'\}}$ , s.t. (a1), and

$x_{uu'} + x_{vu'} \leq 1$, for $\{u,v\} \in \bar{E}$, and $u' \in V'$,

$y_{\{u,v\}\{u',u'\}} \leq x_{uu'}$, $y_{\{u,v\}\{u',u'\}} \leq x_{vu'}$, for $\{u,v\} \in E$, and $u' \in V'$.

If (b1) in Table 2 is substituted for (a1) in the IP formulation, and if $K_{max} = 1$ (i.e., $G'$ has a single node), and letting edge weights be equal to 1, then it solves the *maximum weighted clique* problem (MWC), i.e., the *maximum weighted independent set* problem (MWIS) in the graph $\bar{G}$. Next, we discuss another interesting generalization of graph coloring known as *graph labeling* (GL) with many applications.

*Graph Labeling (GL):*

Input: Positive integers $m, k$ ($m \geq k$) and $\lambda_{mk}$.

Output: Is there an *f(m,k)-labeling* of graph $G=(V,E)$, i.e., is there an assignment $f: V \rightarrow \{0, 1, \ldots, \lambda_{mk}\}$ (coloring of vertices with non-negative integers) such that
 (A)  $|f(u) - f(v)| \geq m$, if $u$ and $v$ are adjacent, and
 (B)  $|f(u) - f(v)| \geq k$, if $u$ and $v$ are distance two apart.

Input: Positive integers $m$ and $k$ ($m \geq k$).

Output: Find an *f(m,k)-labeling* of graph $G=(V,E)$, i.e., $f: V \rightarrow \{0, 1, \ldots, \lambda_{mk}\}$ such that (A) and (B) are satisfied and $\lambda_{mk}$ is the smallest positive integer. The optimal value of $\lambda_{mk}$, denoted by $\lambda_{mk} - number$, called the minimum *span* of $G$.

Applications of graph labeling problems are reported in channel assignment [139], radio labeling [140,141,142], and computer networks [97]. Refer to [1,5,9,143] for comprehensive surveys. In [38] a *natural* IP formulation for the special case of the graph labeling problem is presented. Several generalizations of GL have been presented, [144,145]. These models can be considered as special cases of the *T-coloring* problem [47,48,146]. To present an IP formulation of the GL feasibility problem, we first define a complete graph $G'=(V', E')$ with $V' = \{0, 1, \ldots, \lambda_{mk}\}$ and $E'$ as a set of edges between all pairs $u', v' \in V'$. For each pair $u', v' \in V'$, let $w_{\{u',v'\}} = |u' - v'|$. In graph $G$, for every pair $u, v \in V$ with a distance 2 apart, we add a dummy edge. Denote the set of dummy edges by $E_D$. An IP formulation of the feasibility of GL is now stated in the following. In that, the assignment is between the nodes of two graphs, $\bar{G}=(V, E\cup E_D)$ and $G'=(V', E')$. Feasible and infeasible assignments are similar to the graph coloring problem that appears above.

(a1), and $x_{uu'}+x_{vv'}\le 1$, and $x_{uv'}+x_{vu'}\le 1$, for $\{u,v\}\in E$, and $\{u',v'\}\in E'$, $w_{\{u',v'\}}<m$,

$x_{uu'}+x_{vv'}\le 1$, and $x_{uv'}+x_{vu'}\le 1$, for $\{u,v\}\in E_D$, and $\{u',v'\}\in E'$, $w_{\{u',v'\}}<k$,

$x_{uu'}+x_{vu'}\le 1$, for $\{u,v\}\in E\cup E_D$, and $u'\in V'$.

The formulation can be extended to GL optimization. Let $\lambda_{mk}$ be an upper bound to $\lambda_{mk}-number$. Now, an IP formulation of the problem can be stated as follows.

*Min* $\lambda$, The above feasible conditions, $u'x_{uu'}\le\lambda$, for $u\in V, u'\in V'$, and $\lambda\ge 0$.

Here, the last set of inequalities combined with the minimization of the objective function ensures that the optimal value of $\lambda$ will be equal to $\lambda_{mk}-number$. Next, we discuss the *metric labeling problem* (MLP), which has applications in different areas of engineering, especially in pattern analysis, [147,148,149,150, 151,152]. An IP formulation for special cases was presented in [29,30], and in [150,151] approximation results were presented. Our IP formulation is more general than theirs.

*Metric Labeling Problem (MLP):*

Input: Undirected graph $G=(V, E)$ with a possible self-loop for each node, and weight $w_e$ for each $e\in E$. A set of labels $V'$ with a metric distance function $D: V'\times V'\to R$, and a cost function, $c$, assigning a cost of $c(u,u')$ when two nodes $u\in V$ and $u'\in V'$ are assigned to each other.

Output: Find an assignment, $f: V\to V'$, to minimize P2 (or P4).

In many applications, we have $f(u)\in L_u$ for a given subset $L_u\in V'$. In general, we have $V'=\bigcup_{u\in V}L_u$. If label $u'\in V'$ is assigned to the node $u\in V$, it incurs a cost of $c(u,u')$, and if two labels $u',v'\in V'$ (possibly $u'=v'$) are assigned to the endpoints of an edge $\{u,v\}\in E$, it incurs a cost of $d(\{u,v\},\{u',v'\})=w_{\{u,v\}}D(u',v')$. Define a graph $G'=(V',E')$ where $\{u',v'\}\in E'$ iff $D(u',v')\ne 0$. Originally the MLP was defined as $D(.,.)$ as a metric function [28]. Our formulation does not assume that. We look for a function $f: V\to V'$ to minimize P2 (or P4). An IP formulation that minimizes the objective function P2 can be stated as follows.

*Min* $\sum_{u\in V}\sum_{u'\in L_u}c(u,u')x_{uu'}+\dfrac{1}{2}\sum_{u\in V}\sum_{v\in N(u)}\sum_{u'\in L_u}\sum_{v'\in L_v\cap N[u']}w_{\{u,v\}}D(u',v')y_{\{u,v\}\{u',v'\}}$

s.t. (g), and $\sum_{u'\in L_u}x_{uu'}=1$, for $u\in V$.

An IP formulation of the problem that minimizes the objective function P4 can be stated as follows. Our formulation differs from those in the literature and is more general.

*Min K,* s.t. (g), and $\sum_{u'\in L_u}x_{uu'}=1$, for $u\in V$,

$\sum_{u:u'\in L_u}c(u,u')x_{uu'}+\dfrac{1}{2}\sum_{u:u'\in L_u}\sum_{v\in N(u)}\sum_{v'\in L_v\cap N[u']}w_{\{u,v\}}D(u',v')y_{\{u,v\}\{u',v'\}}\le K$, for $u'\in V'$.

A special case of the above formulation under the *task assignment in distributed computing systems (TADCS)* is discussed in [49] in which a graph-matching formulation for the problem is presented and in which is also some discussion of an integer programming formulation. In the TSDCS, we have a set of tasks, *V,* and a set of processors, *V'*. A *task graph G=(V, E)* represents the set of tasks and their communication links shown by the edges of the graph. A *processor graph G'=(V', E')* represents the set of processors and their communication links shown by the edges of the graph. The problem is to find an assignment function $f: V\to V'$ that assigns the tasks to the processors. If task $u\in V$ is assigned to an eligible processor $u'\in V'$, the processing time is $c(u,u')$. If two tasks $u,v\in V$ need to communicate (i.e., if there is an edge $\{u,v\}\in E$), they can be assigned to one processor; in that case, there is no communication time between them. Two tasks can also be assigned to two different processors if a communication link is possible between them (i.e. if there is an edge $\{u',v'\} \in E'$). In that case, there is a communication time between two tasks measured by $d(\{u,v\},\{f(u),f(v)\})$. The problem with the objective P2 asks for an assignment $f: V \to V'$ to minimize the total times (e.g., [50,51,52,53]). In a problem with objective P4, we want to minimize the completion time of the last task, i.e., a load of the processor with the maximum processing and communication times [49]. Other graph optimization problems related to the MLP with objective function P2 are *Graph Embeddings in d-dimensions, Linear arrangement with d-dimensional cost*, and *Minimizing Storage-Time Product*. The above formulations can be modified for these problems. However, Even, Naor, Rao, and Schieber [28] presented *indirect* IP formulations for each case.

In the next section, we present some variations of graph crossing optimization. We first discuss the *Golomb Ruler problem (GR)* which has a variety of applications including crystallography, radioastronomy, and telecommunication [10,153,154,155].

5.4. Graph Crossing Optimization

A GR with *n* marks is defined as a set of *n* integers, called marks, $0 = a_1 < a_2, \ldots, < a_n$, such that the differences $a_j - a_i$, $1 \le i < j \le n$ are all distinct and nonzero. The optimal GR with *n* marks is a GR with the smallest value of $a_n$, referred to as the *length* of the ruler. The complexity of the problem is still open, [10]. Many good heuristics for large problems are available, [156]. Upper bounds are discussed in [10], and several IP formulations are proposed in [10,40]. We propose a new formulation here.

*Golomb Ruler (GR):*
Input: Two nonnegative integer numbers *n* and *K* with $n \le K$.
Output: Is there a set of *n* integers $\{a_1, \ldots, a_n\} \subseteq \{0, 1, \ldots, K\}$ where $0 = a_1 < a_2, \ldots, < a_n$, such that the differences $a_j - a_i$, $1 \le i < j \le n$ are all distinct?

Input: A nonnegative integer *n*.
Output: A set of *n* integers $0 = a_1 < a_2, \ldots, < a_n$, such that the differences $a_j - a_i$, $1 \le i < j \le n$ are all distinct, and $a_n$ is smallest.

Let $V=\{1,\ldots,n\}$ be a set of *n* integers placed linearly on a line from the smallest to the largest. Similarly, let $V'=\{0,1,\ldots,K\}$ be a set of *K+1* integers placed linearly from the smallest to the largest on a parallel line to the first line. Define a directed graph $G=(V,A)$ where $A=\{(i,j):i<j,$ for $i,j=1,\ldots,n\}$ with weight $w_a = 1$ for $a \in A$, and define a directed graph $G'=(V',A')$ where $A'=\{(i,j):i<j,$ for $i,j=0,1,\ldots,K\}$ with weight $w_{a'} = 1$ for each $a' \in A'$. Consider an assignment $f: V \to V'$, such that each element of *V* is assigned to one element from *V'* and one element of *V'* may be assigned to one element of *V* (the assignment between nodes $u \in V$ and $u' \in V'$ is called an *alignment*). We are looking for an assignment $f: V \to V'$, where no two *alignments* cross, and such that $f(u)-f(v)$ for all $(u,v) \in A$ are distinct. Each element $u \in V$ receives an integer number $a_u = f(u)$ where $a_u < a_{u+1}$ for $u=1,\ldots,n-1$, and differences between these integers are distinct. If we force the smallest element in *V* to be assigned to the first element in *V'*, then we have the desired result for the GR feasibility problem. Below, we state an IP formulation of the feasibility problem.
(a1),(b2),(g), $x_{uv'} + x_{vu'} \le 1$, $1 \le u < v \le n$, and $1 \le u' < v' \le K$,
$\sum_{(u,v) \in A} \sum_{(u',v') \in A', v'-u'=k} y_{(u,v)(u'v')} \le 1$, for *k=1,...,K*, and $x_{11} = 1$.

Here, we used directed graphs to formulate the problem, however, it is easy to modify the formulation for undirected graphs. In the GR optimization problem, it is assumed that an upper bound *K* for the value of $a_n$ is given (e.g., a large enough number). When an element from *V'* is assigned to the largest element of *V*, it carries the length of Golomb Ruler, i.e., $a_n$, that is minimized. The objective function is in the form of P1.
*Min K\**, (a1),(b2),(g), $x_{uv'} + x_{vu'} \le 1$, $1 \le u < v \le n$, and $1 \le u' < v' \le n'$,

$$\sum_{(u,v) \in A} \sum_{(u',v') \in A', v'-u'=k} y_{(u,v)(u',v')} \leq 1, \text{ for } k=1,\ldots,K,$$

$u' x_{uu'} \leq K^*$, for $u' \in V'$, and $x_{11} = 1$.

Next, *Contact Map Problem* (CMP), with applications in computational biology, is discussed (refer to [41,31,91], for recent results). We are given two undirected graphs $G=(V, E)$ and $G'=(V', E')$ where for each set $V$ and $V'$, a pre-specified ordering is given. Each set of vertices with their order represents DNA. An edge in graph $G$ (or $G'$) represents a *contact* if two end vertices are "close enough", measured by a distance measure, with each other. Assume that the vertices of each graph are lined up in the given order on two straight lines parallel to each other. We look for the assignment between the vertices of the two graphs. An assignment between a vertex $u \in V$ and a vertex $u' \in V'$ is called an *alignment*. If the endpoints of an edge are assigned to the endpoints of an edge $\{u',v'\} \in E'$, it is called a *shared contact*. A feasible assignment is obtained when no two *alignments* cross. The objective is to find a feasible assignment between the vertices of two graphs with the maximum number of shared contacts.

*Contact Map Problem (CMP):*

Input: Two undirected graphs $G=(V, E)$ with $|V|=n$, and $G'=(V', E')$ with $|V'|=n'$, and pre-specified ordering for the elements of $V$, and the elements of $V'$.

Output: An assignment $f: V \to V'$ with the maximum number of *shared contacts* where no two alignments are crossed.

The DNA similarity problem was first introduced in graph theoretic format in [94]. It was shown that the objective of CMP is to find the largest subgraph isomorphism between two graphs while preserving the pre-assigned sequence structure of the nodes in each graph and avoiding cross-alignment. Using the formulas in LCS (discussed earlier), we can formulate the problem as follows. In $G$, let $w_e = 1$ for $e \in E$, and in $G'$, let $w_{e'} = 1$ for $e' \in E'$. If two edges $\{u,v\} \in E$ and $\{u',v'\} \in E'$ with $u<v$ and $u'<v'$ are feasibly assigned together, then we must have $x_{uu'} x_{vv'} = 1$. In that, we have $w_e w_{e'} x_{uu'} x_{vv'} = 1$. Now using the LCS results that we discussed earlier an IP formulation of the problem can be stated as follows [41]:

*Max* $\sum_{\{u,v\} \in E, \{u',v'\} \in E'} y_{\{u,v\}\{(u',v')\}}$, s.t. (b1),(b2), $x_{uv'} + x_{vu'} \leq 1$, for $1 \leq u \leq v \leq n$ and $1 \leq u' \leq v' \leq n'$,

$y_{\{u,v\}\{(u',v')\}} \leq x_{uu'}$, $y_{\{u,v\}\{(u',v')\}} \leq x_{vv'}$, for $\{u,v\} \in E$ and $\{u',v'\} \in E'$.

6. CONCLUSION AND REMARKS

In this paper, we introduced general graph optimization approaches. The general problems include a wide range of graph theoretic problems. Natural integer programming formulations for the general problems, as well as a variety of specific matching problems, were presented. Some of these formulations are new and some are alternative to existing ones. The relationship between the general approaches and the quadratic assignment problem was pointed out. Further study should concentrate on applications of the IP formulations in practice. Often real problems have extra constraints that need to be included in the formulations we presented here. Since these problems are NP-hard and in practice problems are large-scale, thus heuristics and meta-heuristics should be designed in connection with formulations presented here as solution procedures. Another line of research may be the use of the formulations presented here to develop lower or upper bounds in algorithmic development for specific problems.

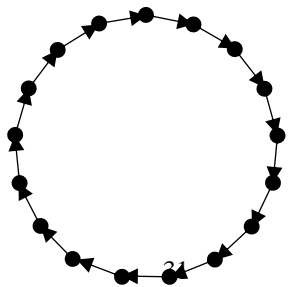

**Fig. 1.** A desired feasible output graph for TSP with k nodes.

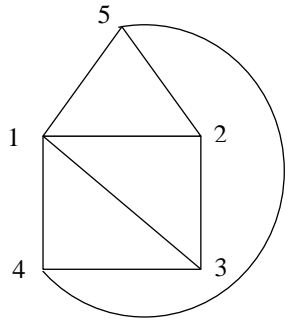

**Fig. 2.** Illustration of subgraph isomorphism (SI) and induced subgraph isomorphism (ISI)

**TABLE 1.** GENERAL OBJECTIVE FUNCTIONS USED IN THE PROBLEMS

$P1: \quad Min_f \left\{ Max_{u \in V, v \in N(u)} \left\{ c(u, f(u)) + d\left(\{u,v\},\{f(u),f(v)\}\right) \right\} \right\},$

$P2: \quad Min_f \left\{ \sum_{u \in V} c(u, f(u)) + \sum_{\{u,v\} \in E} d\left(\{u,v\},\{f(u),f(v)\}\right) \right\},$

$P3: \quad Min_f \left\{ \sum_{u \in V} Max_{v \in N(u)} \left\{ c(u, f(u)) + d\left(\{u,v\},\{f(u),f(v)\}\right) \right\} \right\},$

$P4: \quad Min_f \left\{ Max_{u' \in V'} \left\{ \sum_{u \in V: f(u)=u'} c(u, f(u)) + \sum_{\substack{\{u,v\} \in E: (f(u)=u' \text{ or } f(v)=u') \\ \text{and } \{f(u),f(v)\} \in E'}} d\left(\{u,v\},\{f(u),f(v)\}\right) \right\} \right\},$

$P5: \quad Min_f \left\{ \sum_{u' \in V'} Max_{u \in V: f(u)=u'} \left\{ c(u, f(u)) + \sum_{v \in N(u)} d\left(\{u,v\},\{f(u),f(v)\}\right) \right\} \right\},$

$P6: \quad Min_f \left\{ Max_{u \in V} \left\{ c(u, f(u)) + \sum_{v \in N(u)} d\left(\{u,v\},\{f(u),f(v)\}\right) \right\} \right\},$

$P7: \quad Min_f \left\{ Max_{i' \in V'} \left\{ \sum_{\{u,v\} \in E, f(u) \leq i' < f(v)} \left\{ c(u, f(u)) + c(v, f(v)) + d\left(\{u,v\},\{f(u),f(v)\}\right) \right\} \right\} \right\}.$

*Table 2. Assignment formulas used in formulating problems in part ii*

| | |
|---|---|
| (a1) $\sum_{u' \in V'} x_{uu'} = 1$, for $u \in V$, | (a2) $\sum_{u \in V} x_{uu'} = 1$, for $u' \in V'$, |
| (b1) $\sum_{u' \in V'} x_{uu'} \leq 1$, for $u \in V$, | (b2) $\sum_{u \in V} x_{uu'} \leq 1$, for $u' \in V'$, |
| (c1) $x_{uu'} + x_{vv'} \leq 1$, $x_{uv'} + x_{vu'} \leq 1$, for $\{u,v\} \in E$, and $\{u',v'\} \in \overline{E}'$, | (c2) $x_{u'u} + x_{v'v} \leq 1$, $x_{v'u} + x_{u'v} \leq 1$, for $\{u,v\} \in \overline{E}$, and $\{u',v'\} \in E'$, |
| (d1) $x_{uu'} + x_{vv'} \leq 1$, $x_{uv'} + x_{vu'} \leq 1$, for $(u,v) \in A$, and $(u',v'),(v',u') \notin A'$, | (d2) $x_{uu'} + x_{vv'} \leq 1$, $x_{uv'} + x_{vu'} \leq 1$, for $(u',v') \in A'$, and $(u,v),(v,u) \notin A$ |
| (e) $x_{uv'} + x_{vu'} \leq 1$ for $(u,v) \in A$, and $(u',v') \in A'$, | |
| (f) $x_{uu'} + x_{vv'} \leq y_{\{u,v\}\{u',v'\}} + 1$, $x_{uv'} + x_{vu'} \leq y_{\{u,v\}\{u',v'\}} + 1$, for $\{u,v\} \in E$, and $\{u',v'\} \in E'$, | |
| (g) $x_{uu'} + x_{vv'} \leq y_{(u,v)(u',v')} + 1$, for $(u,v) \in A$, and $(u',v') \in A'$, | |
| (h1) $x_{uu'} + x_{uv'} \leq 1$, for $u \in V$, and $\{u',v'\} \in E'$, | (h2) $x_{uu'} + x_{uv'} \leq 1$, for $u \in V$, and $(u',v') \in A'$, |
| (i1) $x_{uu'} + x_{vu'} \leq 1$, for $\{u,v\} \in E$, and $u' \in V'$, | (i2) $x_{uu'} + x_{vu'} \leq 1$, for $(u,v) \in A$, and $u' \in V'$. |